\documentclass{aa}  
\usepackage{graphicx}
\usepackage{txfonts}
\usepackage{natbib}
\bibpunct{(}{)}{;}{a}{}{,}

\newcommand{\gzinterv}{$0.04-1.51$}
\newcommand{\noim}{48}
\newcommand{\nosys}{21}
\newcommand{\noimwell}{22}
\newcommand{\nosyswell}{13}
\newcommand{\noimbad}{26}

\newcommand{\noimground}{11}

\newcommand{\bvdiff}{0.7}
\newcommand{\perccut}{12}
\newcommand{\ideorv}{2.4}
\newcommand{\ideorvfwhm}{2.7}
\newcommand{\ideorvsigma}{1.1}

\newcommand{\mgfour}{MG0414+0534 }
\newcommand{\sdsstusenfyra}{SDSS J1004+4112}
\newcommand{\hetwo}{HE0230-2130}
\newcommand{\sdssettfyrafyra}{SDSS J144612.98+035154.4}

\newcommand{\cardelli}{CCM}
\newcommand{\fitzpatrick}{Fitzpatrick}
\newcommand{\prevot}{Pr\'{e}vot}
\newcommand{\calzetti}{CAB}

\begin{document}

   \title{Extinction properties of lensing galaxies}

   \author{L. \"{O}stman\inst{1} \and A. Goobar\inst{1} \and
          E. M\"{o}rtsell\inst{2} \fnmsep }

   \offprints{L. \"{O}stman}

   \institute{Department of Physics, Stockholm University, SE 106 91
              Stockholm, Sweden \\ \email{[linda;ariel]@physto.se}
              \and Department of Astronomy, Stockholm University, SE
              106 91 Stockholm, Sweden \\ \email{edvard@astro.su.se} }

   \date{Received -; accepted -}

  \abstract
{Observations of quasars shining through foreground galaxies, offer a
way to probe the dust extinction curves of distant
galaxies. Interesting objects for this study are found in strong
gravitational lensing systems, where the foreground galaxies generate
multiple images.}
%
{The reddening law of lensing galaxies is investigated by studying the
colours of gravitationally-lensed quasars, and a handful of other
quasars where a foreground galaxy is detected.}
%
{We compare the observed colours of quasars reported in the
literature, with spectral templates reddened by different extinction
laws and dust properties. The data consists of {\nosys} quasar-galaxy
systems, with a total of {\noim} images. The galaxies, which are both
early- and late-type, have redshifts in the interval $z=\
${\gzinterv}.}
%
{We measure a difference in rest-frame $B-V$ between the quasar images
we study, and quasars without resolved foreground galaxies. This
difference in colour is indicative of significant dust extinction in
the intervening galaxy. Good fits to standard extinction laws were
found for {\noimwell} of the images, corresponding to {\nosyswell}
different galaxies. Our fits imply a wide range of possible values for
the total-to-selective extinction ratio, $R_V$. The distribution was
found to be broad with a weighted mode of $\bar R_V$={\ideorv} and a
FWHM of $\Delta{R{_V}}$={\ideorvfwhm} ($\sigma_{R_V} \sim$
{\ideorvsigma}). Thus the bulk of the galaxies for which good
reddening fits could be derived, have dust properties compatible with
the Milky Way value ($R_V=3.1$).}
%
{}


   \maketitle
%

\section{Introduction}
\label{sec:intro}

Measuring the reddening due to dust in galaxies, provides information
on dust grains in the interstellar medium. It is in addition important
for dimming corrections, when measuring cosmological distances using
Type Ia supernovae, the most direct probe of the expansion history of
the Universe.

It is often assumed that the wavelength dependence of extinction in
distant galaxies is identical to the average in the Milky
Way. However, there are indications that this is not always the case
\citep[e.g.][]{prevot84,calzetti00}. The purpose of this work is to
examine the reddening relations in galaxies over a wide range of
redshifts, to improve our understanding of galaxy dust properties.

For galaxies in which individual stars are resolved, dust extinction
can be studied using the pair method. In this method, the galaxy
extinction curve is determined by comparing the spectral energy
distribution of reddened and non-reddened stars of the same spectral
type and luminosity class \citep{massa83}. For more distant galaxies,
other methods must be used, such as comparing the galaxy colours with
a dust-free model (although this is subject to large model
dependencies) \citep[see e.g.][]{patil06}, determining the
differential extinction of multiply-imaged quasars
\citep[e.g.][]{falco99}, or measuring the absolute extinction by
foreground galaxies of quasars \citep{ostman06}. 
There are problems using the method of differential extinction for
multiply-imaged quasars. To obtain a unique $R_V$ value, one of the
images must have negligible dust extinction when compared to the
other, or the extinction curves must be similar along the two separate
lines of sight.
The potential problem of this limitation is discussed in
\citet{mcgough05} and \citet{ardis}. If these conditions are not met,
the $R_V$ value derived will be a function of the extinctions along
the two lines of sight, and their $R_V$ values (see Eq.~6 in
\citet{mcgough05}). In this paper, this restriction is circumvented by
studying the individual colours of background quasar images in order
to investigate the reddening by foreground galaxies. This is possible
because of the fairly homogeneous spectral properties of quasars. The
method is very similar to the one used in \citet{ostman06}, where the
focus was on finding systems by matching the positions of quasars with
galaxies. In this paper, the focus is on using systems found through
gravitational lensing.

Studies of colours of Type Ia supernovae have also been used to
constrain the dust properties of their host galaxies
\citep[e.g.][]{joeveer83,branch92,riess96,reindl05,guy05}. When the
reddening correction has been optimised to minimise the scatter of the
Type Ia supernova Hubble diagram, a quite different wavelength
dependence of the extinction curve, when compared to the Milky Way,
has been derived \citep[e.g.][]{2006A&A...447...31A}. It is unclear
whether these findings indicate the presence of non-standard dust
along the line of sight to supernovae, or if they point to a different
source of the brightness-colour relation, perhaps intrinsic to
supernovae themselves. We aim to establish if the reddening laws
derived using supernovae are applicable to different astrophysical
scenarios, such as quasars shining through galaxies.

In Sec.~\ref{sec:general}, different parameterisations of dust
attenuation are discussed. The method for estimating the dust
properties is presented in Sec.~\ref{sec:method}. The results and a
discussion of the results are given in
Sec.~\ref{sec:results}. Finally, we summarise and conclude in
Sec.~\ref{sec:summary}.


\section{Dust extinction parameterisations}
\label{sec:general}

Extinction curves are often parameterised by the reddening parameter
called the total-to-selective extinction ratio $R_V$, which is defined
by
\begin{equation}
R_V \equiv \frac{A_V}{E(B-V)},
\end{equation}
where $A_V$ is the V-band extinction and $E(B-V)$ is the colour excess,
\begin{equation}
E(B-V) \equiv (B-V)_{\rm obs}-(B-V)_{\rm intr}.
\end{equation}
The observed colour is denoted by $(B-V)_{\rm obs}$ and the intrinsic
colour by $(B-V)_{\rm intr}$. The reddening parameter is a measure of
the slope of the extinction curve, and is related to the size of the
dust grains. Large grains produce greyer extinction, with larger
values of $R_V$. 
The average value of $R_V$ in the Milky Way is 3.1, but the value
varies within the interval $2\lesssim R_V \lesssim 6$ between
different regions \citep[see e.g.][and references therein]{draine03}.
The variations are rare and appear to be connected with dense
molecular gas \citep{jenniskens93}. Among distant galaxies, the
variation of $R_V$ could be larger. For example, \citet{falco99}
derived values in the interval $1.5 \lesssim R_V\lesssim 7.2$ using
the method of differential extinction between images of
gravitationally-lensed quasars. Measurements of the mean $R_V$ from
Type Ia supernova samples yield lower values than the Milky Way
average. Some examples are $R_V \sim 1.8$ \citep{2000ApJ...539..658K},
$R_V = 2.55 \pm 0.30$ \citep{riess96}, $R_V = 2.6 \pm 0.4$
\citep{1999AJ....118.1766P}, $R_V = 1.09$ \citep{1998A&A...331..815T},
$R_V = 2.5$ \citep{2004MNRAS.349.1344A}, $R_V=1.75 \pm 0.27$
\citep{2007arXiv0712.1155N} and $\beta=1.77 \pm 0.16$ corresponding to
$R_V = 0.77 \pm 0.16$
\citep{2006A&A...447...31A,2007A&A...466...11G}. Although
parameterised as an interstellar extinction law, it is possible that
the brightness-colour relation observed in Type Ia supernovae, could
be intrinsic to the supernovae or related to circumstellar dust.

\begin{figure}
  \centering
  \includegraphics[width=0.5\textwidth]{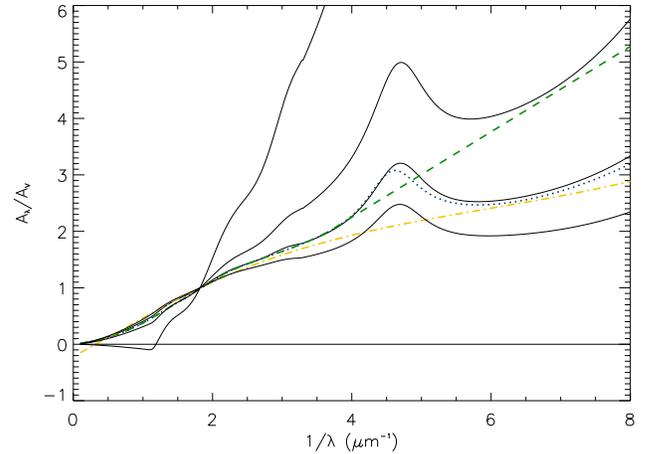}
  \caption{The {\cardelli} parameterisation of the Milky Way
  extinction curve (solid black line) for four different $R_V$ (0.5,
  2, 3.1 and 4) where the highest values of $R_V$ correspond to the
  flattest curves. We also show the {\fitzpatrick} parameterisation
  (dotted blue) with $R_V=3.1$, the {\prevot} law for SMC-like
  extinction (dashed green) with $R_V=3.1$ and the starburst
  extinction law (dashed dotted yellow) with the preferred value for
  starburst galaxies $R_V=4.05$ \citep{calzetti00}. $A_\lambda$ is the
  extinction at wavelength $\lambda$ and $A_V$ is the visual
  extinction.}
  \label{fig:extlaws}
\end{figure}

The extinction curve of galaxies is often described using the
parameterisations by either \citet{cardelli89} (CCM), or by
\citet{fitzpatrick99}. However, some galaxies, such as the Small
Magellanic Cloud (SMC), are not well described by these laws. An
alternative extinction law, valid for the SMC, was presented by
\citet{prevot84}. Starburst galaxies are believed to have a different
extinction law again, which has been described by \citet{calzetti00}
(CAB). All of these extinction laws are shown in
Figure~\ref{fig:extlaws}.

Since the extinction curves for the Small Magellanic Cloud and
starburst galaxies look different compared to the average one for the
Milky Way, the extinction curve of the Milky Way may not be
representative of high-redshift galaxies. Yet, it is often assumed
that the Galactic laws can be applied to dust in other galaxies,
e.g. when correcting the light from supernovae for host galaxy
extinction. In this paper, such an assumption is relaxed.


\section{Fitting dust extinction properties}
\label{sec:method}

We have gathered data for a small number of quasars with foreground
galaxies, from the literature. We require that the redshift of both
the quasar and the galaxy have been determined spectroscopically. We
only consider systems for which apparent magnitudes have been measured
in at least four different filters, which are in the wavelength range
of the redshifted, quasar spectral template. Given these restrictions,
our sample of quasar-galaxy systems consists of {\nosys} foreground
galaxies with a total of {\noim} quasar images. Information about
these pairs are listed in Table~\ref{table:info}.
\begin{table*}
\begin{minipage}[c]{1.0\textwidth}
\renewcommand{\footnoterule}{} 
\caption{Information about the quasar-galaxy pairs in our
sample. $z_Q$ is the redshift of the quasar and $z_G$ is the redshift
of the galaxy. The last two columns list what filters were used in our
analysis and the references for the magnitudes.}
\label{table:info}
\centering
\begin{tabular} {l l c c l l l}
\hline \hline
Quasar & Images & $z_Q$ & $z_G$ & Galaxy type & Filters & Source\footnote{1: {\citet{sdss}}, 2: {\citet{castles}}, 3: {\citet{pindor04}}, 4: {\citet{castander06}}, 5: {\citet{johnston03}}, 6: {\citet{inada03}}, 7: {\citet{toft00}}, 8: {\citet{wisotzki02}}, 9: {\citet{wisotzki99}}, 10: {\citet{morgan04}}, 11: {\citet{oguri04}}} \\
\hline
SDSS J131058.13+010822.2             & A            &  1.39  &  0.04  & star forming & u g r i z        &             1         \\
Q2237+030                            & A B C D      &  1.69  &  0.04  & late & F160W F205W F555W F675W F814W &   2          \\
SDSS J114719.89+522923.1             & A            &  1.99  &  0.05  & starburst & u g r i z                  &  1          \\
SDSS J084957.97+510829.0             & A            &  0.58  &  0.07  & starburst & u g r i z               &      1          \\
SDSS1155+6346                        & A B          &  2.89  &  0.18  & early & F160W F555W F814W K  & 2, 3            \\
MG1654+1346                          & A            &  1.74  &  0.25  & elliptical & F160W F555W F675W F814W& 2            \\
CXOCY J220132.8-320144               & A B          &  3.90  &  0.32  & spiral & g r i Ks                &     4           \\
SDSS J0903+5028                      & A B          &  3.61  &  0.39  & early & F160W F555W F814W r i   & 2, 5               \\
SDSS J0924+0219                      & A B C        &  1.52  &  0.39  & elliptical & u g r i F160W F555W F814W &2, 6            \\
CLASS B1152+199                      & A B          &  1.02  &  0.44  & late & F160W F555W F814W I R V & 2, 7       \\
HE0435-1223                          & A B C D      &  1.69  &  0.45  & S0 & F160W F555W F814W i g r&  2, 8            \\
Q0142-100                            & A B          &  2.72  &  0.49  & early & F160W F555W F675W F814W&  2            \\
HE0230-2130                          & A1 A2 B C    &  2.16  &  0.52  & S0/Sa & B R I K F814W F555W    &  2, 9              \\
BRI0952-0115                         & A B          &  4.43  &  0.63  & elliptical & F160W F555W F675W F814W   &2          \\
WFI2033-4723                         & A1 A2 B C    &  1.66  &  0.66  & Sb/Sc& F160W F555W F814W i   &    2, 10  \\
SDSS J1004+4112                      & A B C D      &  1.74  &  0.68  & cluster of galaxies& u g r i z F160W F555W F814W & 2, 11     \\
J1004+1229                           & A B          &  2.65  &  0.95  & elliptical & F160W F555W F814W I         &  2  \\
MG0414+0534                          & A1 A2 B C    &  2.64  &  0.96  & early & F160W F110W F205W F675W F814W & 2 \\
SDSS J012147.73+002718.7             & A            &  2.22  &  1.39  &unknown& u g r i z     &       1 \\
SDSS J145907.19+002401.2             & A            &  3.01  &  1.39  &unknown& u g r i z      &      1 \\
SDSS J144612.98+035154.4             & A            &  1.95  &  1.51  &unknown& u g r i z     &       1 \\
\hline
\end{tabular}
\end{minipage}
\end{table*}
Most of these are found from strong lensing surveys. However, three of
them comes from analysing the quasar spectra \citep{wang04} and an
additional three are found from coordinate matching \citep{ostman06}.

\begin{figure}
  \centering
  \includegraphics[width=0.5\textwidth]{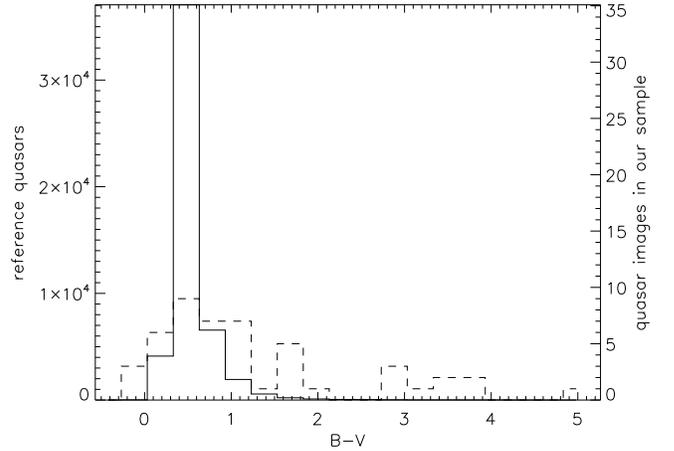}
  \caption{A histogram over rest-frame $B-V$ for the quasars in SDSS
  DR3 (solid line), and for the quasar images in our sample (dashed
  line).}
  \label{fig:bv}
\end{figure}

To test whether our sample of {\nosys} quasars with foreground
galaxies differs in the colours from the main bulk of quasars, we
estimate the rest-frame $B-V$ colour, by K-correcting from the
observed bands, for our set of quasars and compare it with a reference
set from release three of the Sloan Digital Sky Survey (SDSS)
\citep{schneider05}. Table~\ref{table:bv} contains the estimated
values of $B-V$ for our images, and Figure~\ref{fig:bv} shows the
histogram of the two distributions.
\begin{table}
\caption{Estimated rest-frame $B-V$ colour for the different quasar
images in our sample.}
\label{table:bv}
\centering
\begin{tabular} {l l l}
\hline \hline
Quasar & Images & $B-V$ \\
\hline
SDSS J131058.13+010822.2& A& 0.8\\
Q2237+030& A B C D& 3.4,  2.1,  3.6,  3.4\\
SDSS J084957.97+510829.0& A& 1.1\\
SDSS1155+6346& A B& 1.5,  1.1\\
MG1654+1346& A& 0.4\\
CXOCY J220132.8-320144& A B& 1.7,  1.8\\
SDSS J0903+5028& A B& 0.7,  0.5\\
SDSS J0924+0219& A B C& 0.4,  0.6,  0.2\\
CLASS B1152+199& A B& 1.8,  1.7\\
HE0435-1223& A B C D& 1.1,  1.0,  1.0,  1.0\\
Q0142-100& A B& 0.1,  0.1\\
HE0230-2130& A1 A2 B C& 0.4,  0.2,  0.3,  0.8\\
BRI0952-0115& A B& 0.7,  0.8\\
WFI2033-4723& A1 A2 B C& 0.4, -0.1, -0.3, -0.1\\
SDSS J1004+4112& A B C D& 0.1,  0.6,  0.3,  1.6\\
J1004+1229& A B& 3.9,  5.0\\
MG0414+0534& A1 A2 B C& 2.9,  3.1,  2.8,  2.8\\
SDSS J012147.73+002718.7& A& 1.0\\
SDSS J145907.19+002401.2& A& 0.9\\
SDSS J144612.98+035154.4& A& 0.8\\
\hline
\end{tabular}
\end{table}
We find that our set is redder than the comparison set. The difference
in mean rest-frame $B-V$ between the two sets is {\bvdiff}
magnitudes. There is also one quasar, WFI2033-4723, for which three
out of its four images are significantly bluer than the reference set.
The two distributions of rest-frame colours were compared using the
Kolmogorov-Smirnov test, yielding a negligible probability that the
difference is only due to statistical fluctuations. We conclude that
there must be quasars in our sample whose colours are affected by
reddening in the intervening galaxy.

The quasar spectral template used, is a combination of the HST
radio-quiet composite spectrum \citep{telfer02}, and the SDSS median
composite spectrum \citep{vandenberk01}. To simulate the effects of
dust attenuation, the template spectrum is redshifted and reddened, by
different values of $R_V$ and $E(B-V)$. This enables synthetic colours
to be estimated for dust-reddened quasars. To determine the values of
the dust parameters that best describe the observed magnitudes, we
compare the observed colours with the synthetic colours for different
$R_V$ and $E(B-V)$. When choosing colours $X=i-j$ to compare among all
possible filter combinations, we want to use the colour combinations
that have the greatest potential to separate cases of reddening from
those of no reddening. We select the combinations that provide a large
difference between observed colour and expected colour without dust
extinction, $E(i-j)$, and have small errors. We select colours by
maximising the sum over $Q_{ij}$, which is defined by
\begin{equation}
Q_{ij} = \frac{\left[E(i-j)\right]^2}{\sigma_{ij}^2},
\end{equation}
where $\sigma_{ij}$ is the total error of the colour difference, i.e.
the errors of the observed magnitudes combined with the uncertainty in
the template colours. The sum over $Q_{ij}$ runs over $f-1$ colours
where $f$ is the number of filters, and each filter must be used at
least once. The colours chosen from maximising the sum over $Q_{ij}$
are used when calculating the $\chi^2$ described below. For example,
when observations are available in the $ugri$ filter set, the six
possible colour combinations are $u-g$, $u-r$, $u-i$, $g-r$, $g-i$ and
$r-i$. We assume that $Q_{ug}$ and $Q_{ui}$ have the largest values of
$Q_{ij}$. From the remaining four combinations, $u-r$, $g-r$, $g-i$
and $r-i$, we can only choose from $u-r$, $g-r$, and $r-i$, because
the colour $u-i$ can be constructed by combining the colours we have
already picked, $u-g$ and $u-i$. We then choose the filter
combination, of the three remaining, with the largest $Q_{ij}$ to
obtain the three colours to be used in the analysis.

\begin{figure}
  \centering
  \includegraphics[width=0.5\textwidth]{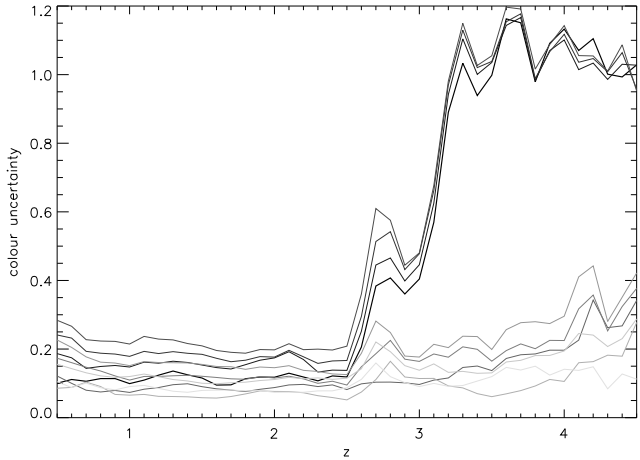}
  \caption{Uncertainties in the template colours for $u-g$, $u-r$,
  $u-i$, $u-z$, $g-r$, $g-i$, $g-z$, $r-i$, $r-z$, $i-z$ (from black
  to light grey) as a function of redshift $z$.}
  \label{fig:intrerror}
\end{figure}

For each quasar image, the best-fit dust extinction is determined by
comparing the synthetic colours, $X_{\rm syn} = X_{\rm
syn}[R_V,E(B-V)]$, with the observed ones, $X_{\rm obs}$, and
minimising the $\chi^2=\chi^2[R_V,E(B-V)]$, which is defined to be
\begin{equation}
 \chi^2 = (X_{\rm syn}-X_{\rm obs})^T{\cal V}^{-1}(X_{\rm syn}-X_{\rm obs}).
\end{equation}
${\cal V}$ is the covariance matrix including the errors from the
observed and synthetic colours. The errors of the synthetic colours
account for both the intrinsic variability of quasars, and the
possible dust extinction in the host galaxy of the quasar. These
errors are calculated from a set of quasars with similar redshifts
($|\Delta z| < 0.05$), without any resolved foreground galaxies. These
were taken from the third data release of the Sloan Digital Sky Survey
\citep{schneider05}. Their colours were $K$-corrected from the SDSS
filters, to the filters used in the observations considered in this
work. To avoid a large impact from colour outliers, quasars with a
colour differing by more than three standard deviations from the mean
were excluded. About {\perccut}\% of the quasars were rejected by this
cut.  In Figure~\ref{fig:intrerror}, the uncertainties for the colours
from the template are shown for the SDSS filters at different
redshifts.
As can be seen in the figure, all colours involving the $u$ band have
very large errors for quasars with $z \gtrsim 2.5$. It is also at $z
\sim 2.5$ that the number of available comparison quasars falls
drastically. The covariance matrix also includes the correlations
between estimated magnitudes in different filters and consideration to
the fact that the same filter can appear more than once in a set of
colour comparisons.

The quasar images were fitted with both the {\cardelli} law and the
{\fitzpatrick} law with $R_V$ and $E(B-V)$ as free parameters. We also
considered SMC like extinction, according to the {\prevot} law. The
{\cardelli} law that we use is the improved version of the relation by
\citet{cardelli89}, where the extinction law, for the wavelength range
1.1 $\mu$m$^{-1}$ $\le 1/\lambda \le$ 3.3 $\mu$m$^{-1}$, has been
exchanged for the curve provided by \citet{odonnell94}. The {\prevot}
law used here consists of two parts. For $\lambda<2700$ {\AA}, we use
a linear fit to the SMC data in \citet{prevot84}, and at longer
wavelengths we use the extinction law by \citet{fitzpatrick99}.

In addition to dust in the foreground galaxy, we have also
investigated the possibility of extinction in the host galaxy of the
quasar.


\begin{table*}
\caption{Best fits of $R_V$ and $E(B-V)$ together with the 1$\sigma$
uncertainties. Note that the lowest possible value for $R_V$ here is
0.1, since $R_V=0$ usually is not defined. Also note that for some
redshifts and filter combinations, there is no $R_V$ dependence for
the {\prevot} law. The interval is then given as $0.1-6.0$. $z_d$ is
the redshift of the dust extinction, which is $z_G$ for foreground
extinction and $z_Q$ for host extinction. $d$ is the impact parameter
expressed in kpc. When this was unavailable, it is marked with a
hyphen in the table. When several laws and/or dust redshifts is either
fitted for one image, they are all listed in the table below. The
probability for each to obtain a higher $\chi^2$ value than what was
obtained for the fit is given in the last column. We include all fits
with a probability larger than 0.5\%. The different images are
separated by a horizontal line.}
\label{tab:fits}
\centering
\begin{tabular} {l c c c c c c c c c}
\hline \hline
Quasar & Image & $z_d$ & d (kpc) & $R_V$ & $E(B-V)$ & Law & Prob \\
\hline
SDSS J131058.13+010822.2  &A      & 0.04  & 15.0  &3.5 (1.4-6.0)&0.3 (0.2-0.4)&CCM              &   96 \% \\
SDSS J131058.13+010822.2  &A      & 0.04  & 15.0  &3.5 (1.8-6.0)&0.3 (0.2-0.4)&Fitzpatrick      &   94 \% \\
SDSS J131058.13+010822.2  &A      & 0.04  & 15.0  &3.5 (1.8-6.0)&0.3 (0.2-0.4)&SMC              &   95 \% \\
SDSS J131058.13+010822.2  &A      & 1.39  & 15.0  &0.1 (0.1-2.5)&0.1 (0.1-0.2)&Host CCM         &   12 \% \\
SDSS J131058.13+010822.2  &A      & 1.39  & 15.0  &1.1 (0.3-2.2)&0.1 (0.0-0.2)&Host Fitzpatrick  &   46 \% \\
SDSS J131058.13+010822.2  &A      & 1.39  & 15.0  &0.1 (0.1-6.0)&0.1 (0.1-0.2)&Host SMC         &   20 \% \\
\hline
Q2237+030                 &A      & 0.04  &  0.7  &0.1 (0.1-0.2)&0.7 (0.5-0.9)&Fitzpatrick      &    1 \% \\
Q2237+030                 &A      & 0.04  &  0.7  &0.1 (0.1-0.2)&0.7 (0.5-0.8)&SMC              &    2 \% \\
\hline
Q2237+030                 &C      & 0.04  &  0.6  &0.1 (0.1-0.2)&1.2 (0.9-1.4)&CCM              &   11 \% \\
Q2237+030                 &C      & 0.04  &  0.6  &0.2 (0.1-0.6)&0.8 (0.6-1.0)&Fitzpatrick      &    4 \% \\
Q2237+030                 &C      & 0.04  &  0.6  &0.2 (0.1-0.6)&0.8 (0.6-1.0)&SMC              &    4 \% \\
\hline
Q2237+030                 &D      & 0.04  &  0.7  &0.3 (0.1-0.6)&0.7 (0.5-0.9)&CCM              &   56 \% \\
Q2237+030                 &D      & 0.04  &  0.7  &0.9 (0.6-1.3)&0.5 (0.3-0.7)&Fitzpatrick      &   23 \% \\
Q2237+030                 &D      & 0.04  &  0.7  &0.9 (0.6-1.3)&0.5 (0.3-0.7)&SMC              &   23 \% \\
\hline
SDSS J114719.89+522923.1  &A      & 0.05  &    -  &0.1 (0.1-3.0)&0.2 (0.1-0.4)&CCM              &   22 \% \\
SDSS J114719.89+522923.1  &A      & 0.05  &    -  &0.3 (0.1-1.6)&0.2 (0.0-0.3)&Fitzpatrick      &   65 \% \\
SDSS J114719.89+522923.1  &A      & 0.05  &    -  &1.0 (0.1-5.4)&0.2 (0.1-0.4)&SMC              &    6 \% \\
SDSS J114719.89+522923.1  &A      & 1.99  &    -  &0.1 (0.1-3.3)&0.0 (0.0-0.1)&Host CCM         &    5 \% \\
SDSS J114719.89+522923.1  &A      & 1.99  &    -  &0.6 (0.4-2.4)&0.0 (0.0-0.1)&Host Fitzpatrick  &   18 \% \\
SDSS J114719.89+522923.1  &A      & 1.99  &    -  &2.5 (0.1-6.0)&0.1 (0.0-0.1)&Host SMC         &   14 \% \\
\hline
SDSS J084957.97+510829.0  &A      & 0.07  & 20.0  &2.3 (0.9-3.6)&0.5 (0.4-0.7)&CCM              &   67 \% \\
SDSS J084957.97+510829.0  &A      & 0.07  & 20.0  &2.2 (1.2-3.2)&0.5 (0.4-0.7)&Fitzpatrick      &   88 \% \\
SDSS J084957.97+510829.0  &A      & 0.07  & 20.0  &2.2 (0.9-3.5)&0.6 (0.4-0.7)&SMC              &   62 \% \\
SDSS J084957.97+510829.0  &A      & 0.58  & 20.0  &0.1 (0.1-6.0)&0.3 (0.2-0.4)&Host SMC         &    2 \% \\
\hline
SDSS1155+6346             &B      & 2.89  &  0.6  &6.0 (0.1-6.0)&0.1 (0.0-0.2)&Host CCM         &    1 \% \\
\hline
MG1654+1346               &A      & 0.25  & 11.3  &0.1 (0.1-0.5)&0.2 (0.0-0.3)&SMC              &    1 \% \\
\hline
CXOCY J220132.8-320144    &A      & 0.32  &    -  &2.2 (1.5-3.1)&0.8 (0.7-1.0)&CCM              &    1 \% \\
CXOCY J220132.8-320144    &A      & 0.32  &    -  &2.4 (1.8-3.3)&0.8 (0.6-1.0)&Fitzpatrick      &    1 \% \\
CXOCY J220132.8-320144    &A      & 0.32  &    -  &2.5 (1.9-3.5)&0.8 (0.6-1.0)&SMC              &    1 \% \\
CXOCY J220132.8-320144    &A      & 3.90  &    -  &0.6 (0.1-3.0)&0.1 (0.1-0.2)&Host CCM         &    3 \% \\
CXOCY J220132.8-320144    &A      & 3.90  &    -  &1.1 (0.2-2.3)&0.1 (0.0-0.2)&Host Fitzpatrick  &   25 \% \\
CXOCY J220132.8-320144    &A      & 3.90  &    -  &6.0 (0.1-6.0)&0.2 (0.2-0.2)&Host SMC         &    8 \% \\
\hline
CXOCY J220132.8-320144    &B      & 0.32  &    -  &2.2 (1.4-3.2)&0.8 (0.6-0.9)&CCM              &    1 \% \\
CXOCY J220132.8-320144    &B      & 3.90  &    -  &0.1 (0.1-3.7)&0.1 (0.1-0.3)&Host CCM         &    8 \% \\
CXOCY J220132.8-320144    &B      & 3.90  &    -  &1.4 (0.4-3.0)&0.1 (0.0-0.2)&Host Fitzpatrick  &   26 \% \\
CXOCY J220132.8-320144    &B      & 3.90  &    -  &6.0 (0.1-6.0)&0.2 (0.1-0.2)&Host SMC         &    4 \% \\
\hline
SDSS J0903+5028           &A      & 3.61  & 10.5  &2.2 (0.1-3.0)&0.2 (0.1-0.3)&Host CCM         &   82 \% \\
SDSS J0903+5028           &A      & 3.61  & 10.5  &2.9 (2.5-3.2)&0.3 (0.2-0.3)&Host Fitzpatrick  &   81 \% \\
\hline
SDSS J0903+5028           &B      & 0.39  &  3.4  &2.3 (1.1-4.5)&0.6 (0.4-0.8)&CCM              &   92 \% \\
SDSS J0903+5028           &B      & 0.39  &  3.4  &2.8 (1.8-4.3)&0.5 (0.3-0.7)&Fitzpatrick      &   92 \% \\
SDSS J0903+5028           &B      & 0.39  &  3.4  &2.8 (1.8-4.4)&0.5 (0.3-0.7)&SMC              &   91 \% \\
SDSS J0903+5028           &B      & 3.61  &  3.4  &0.2 (0.1-4.1)&0.1 (0.1-0.4)&Host CCM         &   45 \% \\
SDSS J0903+5028           &B      & 3.61  &  3.4  &0.6 (0.2-3.4)&0.0 (0.0-0.3)&Host Fitzpatrick  &   92 \% \\
SDSS J0903+5028           &B      & 3.61  &  3.4  &0.1 (0.1-6.0)&0.1 (0.1-0.2)&Host SMC         &   96 \% \\
\hline
Q0142-100                 &A      & 0.49  & 10.9  &6.0 (2.5-6.0)&0.1 (0.0-0.2)&CCM              &   15 \% \\
Q0142-100                 &A      & 0.49  & 10.9  &6.0 (2.7-6.0)&0.1 (0.0-0.2)&Fitzpatrick      &   15 \% \\
Q0142-100                 &A      & 0.49  & 10.9  &6.0 (2.5-6.0)&0.1 (0.0-0.2)&SMC              &   14 \% \\
Q0142-100                 &A      & 2.72  & 10.9  &6.0 (0.1-6.0)&0.1 (0.0-0.2)&Host CCM         &   61 \% \\
Q0142-100                 &A      & 2.72  & 10.9  &6.0 (1.7-6.0)&0.1 (0.0-0.1)&Host Fitzpatrick  &   48 \% \\
Q0142-100                 &A      & 2.72  & 10.9  &0.1 (0.1-6.0)&0.0 (0.0-0.1)&Host SMC         &    2 \% \\
\hline
Q0142-100                 &B      & 0.49  &  2.2  & -  &0.0 (0.0-0.1)& all             &   21 \% \\
\hline
BRI0952-0115              &A      & 0.63  &  4.3  &0.1 (0.1-3.4)&0.2 (0.1-0.3)&CCM              &   18 \% \\
BRI0952-0115              &A      & 0.63  &  4.3  &0.5 (0.2-2.0)&0.2 (0.1-0.4)&Fitzpatrick      &   33 \% \\
BRI0952-0115              &A      & 0.63  &  4.3  &0.1 (0.1-1.6)&0.4 (0.1-0.6)&SMC              &   35 \% \\
BRI0952-0115              &A      & 4.43  &  4.3  &4.2 (0.1-5.4)&0.1 (0.0-0.4)&Host CCM         &   26 \% \\
BRI0952-0115              &A      & 4.43  &  4.3  &4.0 (1.0-5.7)&0.1 (0.0-0.3)&Host Fitzpatrick  &   27 \% \\
BRI0952-0115              &A      & 4.43  &  4.3  &0.1 (0.1-6.0)&0.0 (0.0-0.0)&Host SMC         &    6 \% \\
\hline
\end{tabular}
\end{table*}
\begin{table*}
\centering
\begin{tabular} {l c c c c c c c c c}
\hline \hline
Quasar & Image & $z_d$ & d (kpc) & $R_V$ & $E(B-V)$ & Law & Prob \\
\hline
BRI0952-0115              &B      & 0.63  &  2.4  &0.1 (0.1-3.7)&0.2 (0.1-0.3)&CCM              &   33 \% \\
BRI0952-0115              &B      & 0.63  &  2.4  &0.5 (0.2-2.1)&0.2 (0.1-0.4)&Fitzpatrick      &   65 \% \\
BRI0952-0115              &B      & 0.63  &  2.4  &0.1 (0.1-2.7)&0.4 (0.1-0.6)&SMC              &   42 \% \\
BRI0952-0115              &B      & 4.43  &  2.4  &4.1 (0.1-5.4)&0.1 (0.0-0.4)&Host CCM         &   53 \% \\
BRI0952-0115              &B      & 4.43  &  2.4  &3.8 (0.9-5.7)&0.1 (0.0-0.3)&Host Fitzpatrick  &   55 \% \\
BRI0952-0115              &B      & 4.43  &  2.4  &0.1 (0.1-6.0)&0.0 (0.0-0.0)&Host SMC         &    9 \% \\
\hline
MG0414+0534               &A1     & 0.96  &  9.7  &1.5 (1.4-1.6)&1.7 (1.5-1.8)&Fitzpatrick      &   16 \% \\
MG0414+0534               &A1     & 0.96  &  9.7  &1.3 (1.2-1.5)&1.8 (1.7-2.0)&SMC              &   58 \% \\
\hline
MG0414+0534               &A2     & 0.96  &  9.5  &1.5 (1.5-1.7)&1.9 (1.8-2.1)&Fitzpatrick      &    3 \% \\
MG0414+0534               &A2     & 0.96  &  9.5  &1.4 (1.3-1.5)&2.1 (1.9-2.3)&SMC              &   18 \% \\
\hline
MG0414+0534               &B      & 0.96  & 10.5  &1.5 (1.3-1.6)&1.6 (1.4-1.7)&Fitzpatrick      &   19 \% \\
MG0414+0534               &B      & 0.96  & 10.5  &1.2 (1.1-1.5)&1.8 (1.6-1.9)&SMC              &   57 \% \\
\hline
MG0414+0534               &C      & 0.96  &  7.3  &1.4 (1.3-1.5)&1.5 (1.4-1.7)&Fitzpatrick      &    6 \% \\
MG0414+0534               &C      & 0.96  &  7.3  &1.2 (1.1-1.4)&1.8 (1.6-1.9)&SMC              &   64 \% \\
\hline
SDSS J012147.73+002718.7  &A      & 1.39  &    -  &0.1 (0.1-3.8)&0.1 (0.1-0.2)&CCM              &   58 \% \\
SDSS J012147.73+002718.7  &A      & 1.39  &    -  &2.0 (1.1-3.2)&0.2 (0.1-0.2)&Fitzpatrick      &   70 \% \\
\hline
SDSS J145907.19+002401.2  &A      & 1.39  &    -  &0.1 (0.1-3.8)&0.1 (0.1-0.2)&CCM              &    3 \% \\
SDSS J145907.19+002401.2  &A      & 1.39  &    -  &2.2 (0.2-5.9)&0.2 (0.0-0.4)&Fitzpatrick      &    5 \% \\
SDSS J145907.19+002401.2  &A      & 1.39  &    -  &0.1 (0.1-6.0)&0.2 (0.1-0.2)&SMC              &    2 \% \\
SDSS J145907.19+002401.2  &A      & 3.01  &    -  &0.3 (0.3-0.8)&0.0 (0.0-0.0)&Host Fitzpatrick  &    5 \% \\
SDSS J145907.19+002401.2  &A      & 3.01  &    -  &3.0 (0.1-6.0)&0.1 (0.1-0.1)&Host SMC         &   12 \% \\
\hline
\end{tabular}
\end{table*}


\section{Results and discussion}
\label{sec:results}

Of the {\noim} images, {\noimbad} yielded inconclusive results. The
chi-square of their fits was so high that, if they had been affected
by dust following our extinction laws, there was only a 0.5\%
probability that such a high chi-square value would be obtained.
However, a poor fit does not automatically imply that there is a
strange, unknown reddening law. It can also be the result of for
example (1) underestimated observational errors, (2) the extinction
law having more degrees of freedom than the laws we have assumed, (3)
peculiar quasars that have an intrinsic spectrum that differs
significantly from the template, (4) effects from other intervening
objects that are unresolved in the available images, (5) microlensing
effects, or (6) changes in colour due to the time variability of the
quasar. The significance of the latter two possibilities, when
comparing quasar images, are discussed in \citet{yonehara07}.

\begin{figure*}
  \centering
  \includegraphics[width=0.4\textwidth]{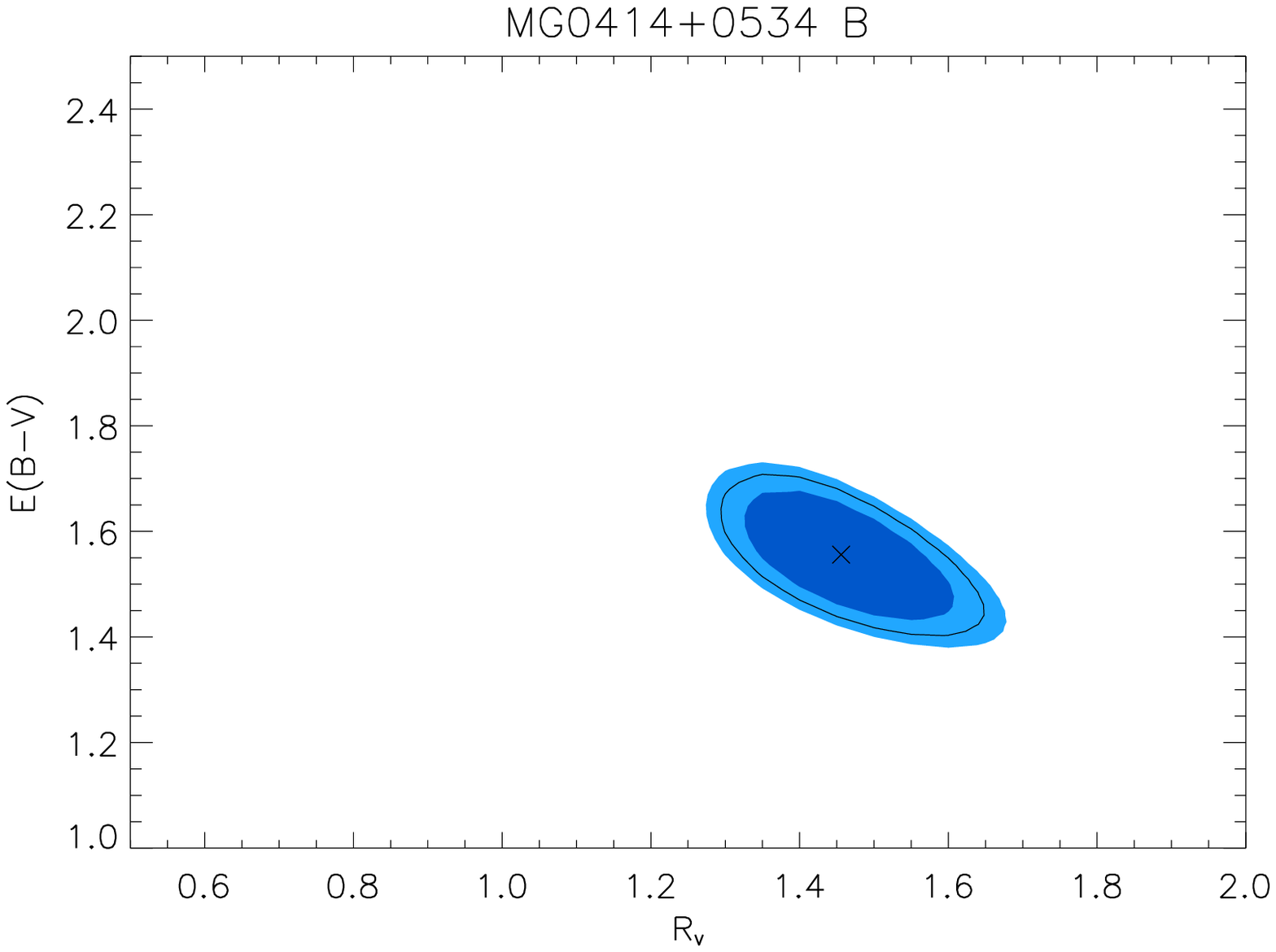}
  \includegraphics[width=0.4\textwidth]{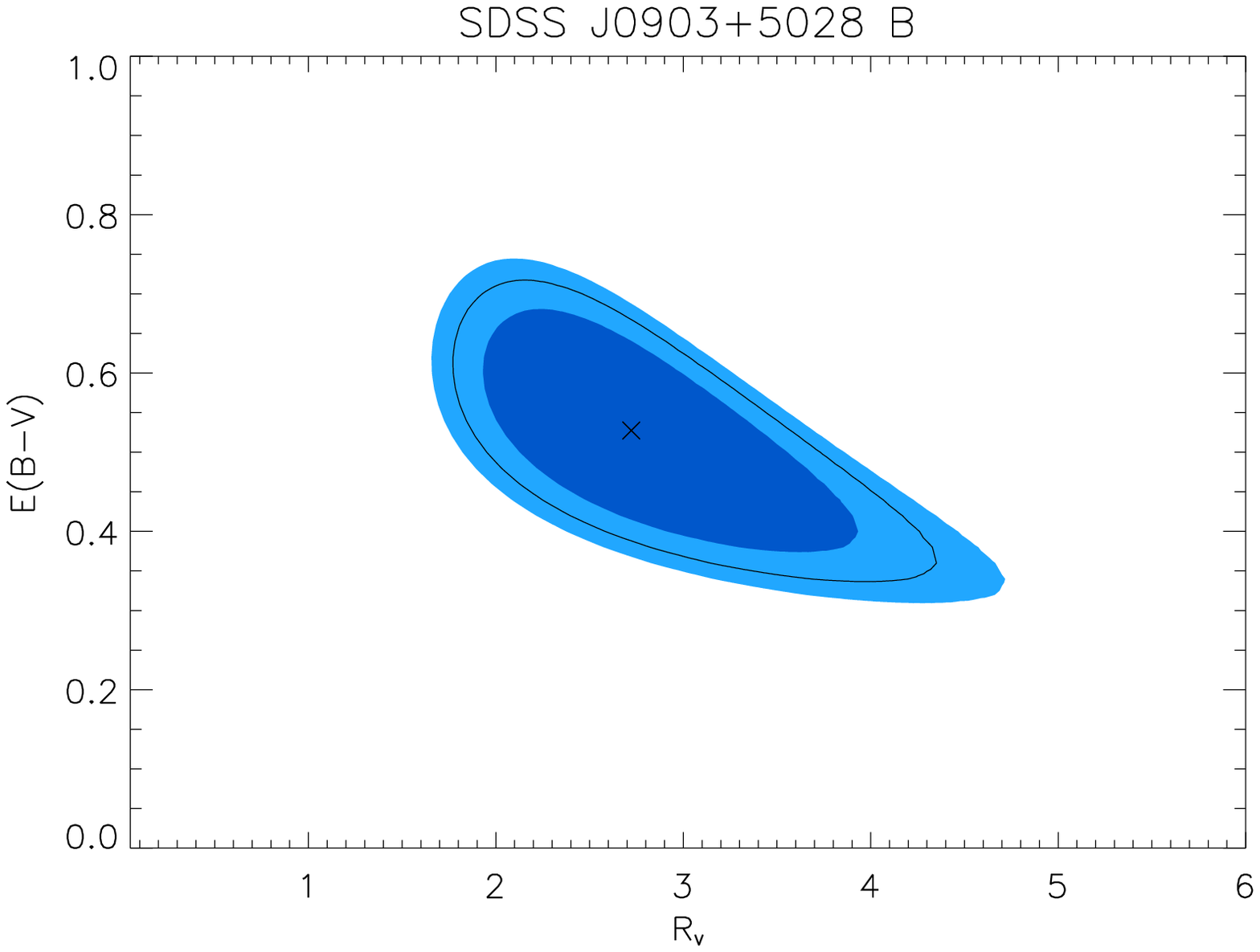}
  \includegraphics[width=0.4\textwidth]{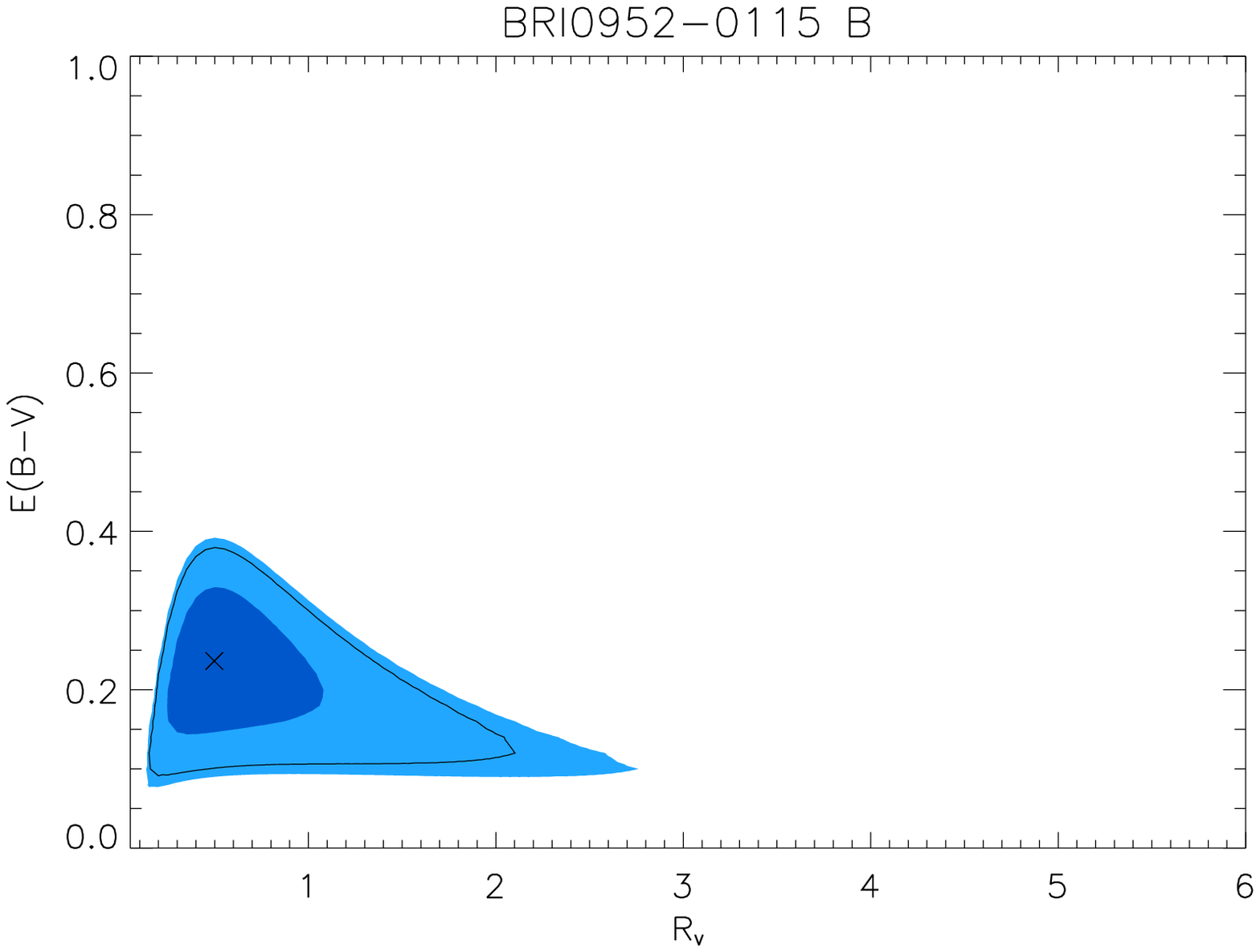}
  \includegraphics[width=0.4\textwidth]{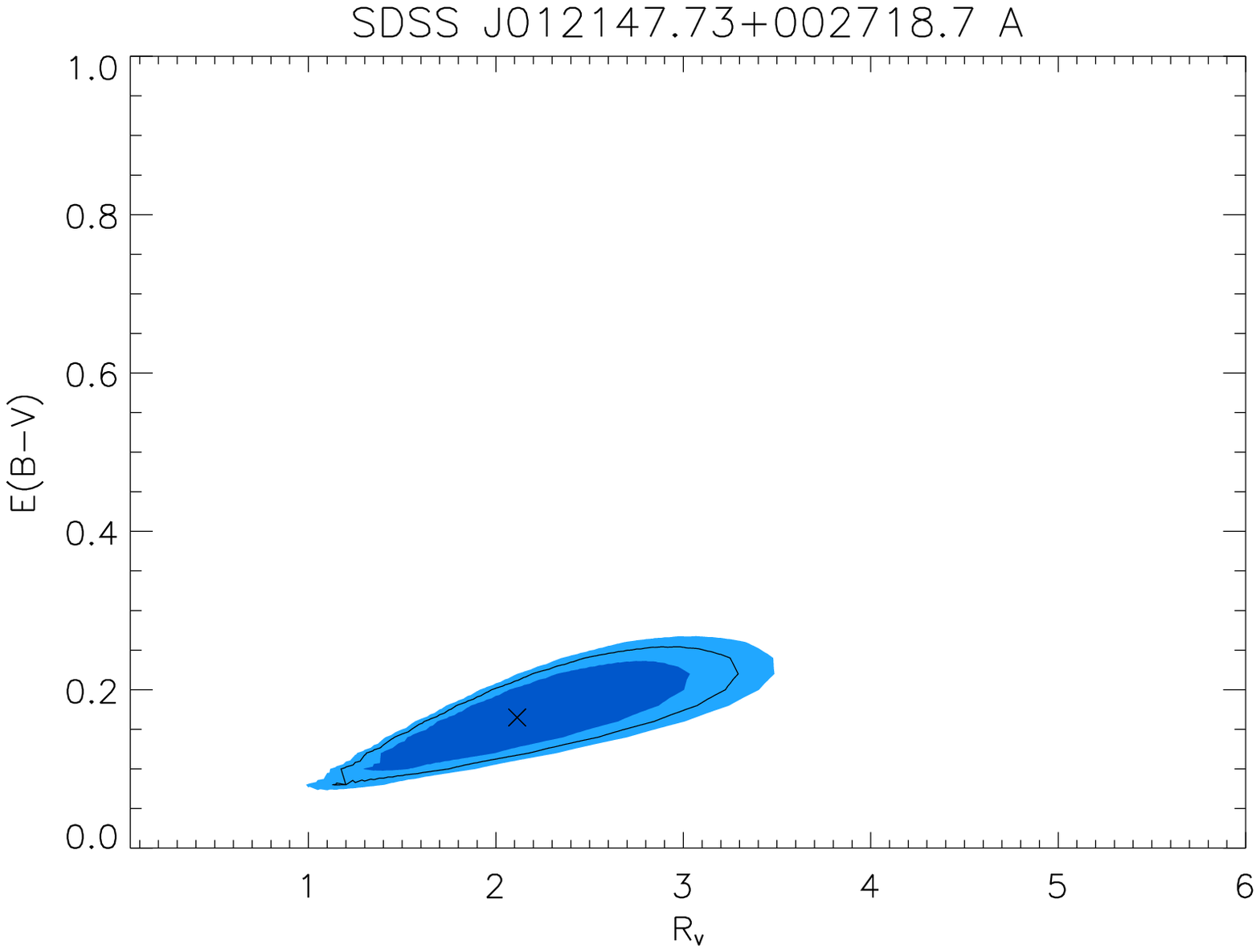}
  \caption{Confidence levels for some of the quasar images that could
  be fitted by dust extinction in the intervening galaxy using the
  {\fitzpatrick} extinction law. The levels correspond to 1$\sigma$
  for one parameter (black line) and 68\% (dark blue region) and 90\%
  (pale blue) for two parameters.}
  \label{fig:contours}
\end{figure*}

The fit results for the remaining {\noimwell} images (corresponding to
{\nosyswell} galaxies) are presented in Table~\ref{tab:fits}. In
Figure~\ref{fig:contours}, a few of the confidence levels obtained
from the $\chi^2$ fitting of the quasar images are shown.
The error intervals in Table~\ref{tab:fits} indicate the $1\sigma$
uncertainty. The errors are calculated with the assumption that the
extinction law is valid for a positive colour excess, and for all
positive values of $R_V$ up to 6.0. Negative values of $E(B-V)$ are
not expected for normal quasars, affected by dust. However, an
intrinsically blue quasar is best fitted with a negative $E(B-V)$.
For small and negative values of $R_V$ ($\lesssim 0.7$), the
extinction laws have a peculiar behaviour (see e.g. the {\cardelli}
parameterisation with $R_V=0.5$ in Figure~\ref{fig:extlaws}). The
non-smooth function, and the negative values of the extinction
$A_\lambda/A_V$, are probably hard to explain using a physical dust
model.
We chose however to consider all fits, down to $R_V = 0$. For large
(positive) values of $R_V$, the extinction becomes less
wavelength-dependent, and it becomes more difficult to distinguish
from the case of no extinction. When the colours of a quasar image in
our analysis can be explained without dust, there is no limitation on
$R_V$, because there is no $R_V$ dependence in the extinction laws for
$E(B-V)=0$.

The law which we refer to as the {\prevot} law, consists of two
parts. For $\lambda<2700$ {\AA}, we use a linear fit to the SMC data
in \citet{prevot84}, and for $\lambda>2700$ {\AA}, we use the
parameterisation of the Milky Way extinction by
\citet{fitzpatrick99}. If the foreground galaxy, where the extinction
occurs, is located at low redshift, most of the filters used for
observations will correspond to regions of the spectrum that are
affected by the second part of the extinction curve, which is
identical to the {\fitzpatrick} law ($\lambda>2700$ {\AA}). If, on the
other hand, the extinction redshift is high, the filters will
correspond to regions of the spectrum that are affected by the first
part of the extinction curve. This part has the relation
$A_\lambda/E(B-V) = a + bx + R_V$. It is therefore not possible to
estimate $R_V$ from the observed colour, e.g. $B-V$, because the $R_V$
dependence is subtracted out, $B-V = A_B-A_V = E(B-V)(a + bx_B +
R_V)-E(B-V)(a + bx_V + R_V) = E(B-V) (bx_B-bx_V)$. This is not the
case for e.g. the {\fitzpatrick} or {\cardelli} law. For objects which
have high extinction redshifts, no preferred value of $R_V$ can be
determined for SMC-like extinction, using this parameterisation. The
same problem occurs for the {\calzetti} law, where a preferred value
of $R_V$ cannot be determined for any extinction redshift, because the
extinction law provides colours that are independent of $R_V$. This is
because the parameterisation we use is $A_\lambda/E(B-V) = a + bx +
cx^2 + dx^3 + R_V$, and thus $B-V = A_B-A_V$ is not dependent on
$R_V$. This should be remembered when studying the fitted values in
Table~\ref{tab:fits}.

We note that even though we have considered both intervening and host
extinction, only host extinction cannot explain the difference in
$B-V$ that we see, when comparing the colours of our quasars with
quasars without intervening galaxies (see
Figure~\ref{fig:bv}). Furthermore, we have taken the possibility of
dust extinction within the host into account, in the calculation of
the covariance matrix.

\begin{figure*}
  \centering
  \includegraphics[bb=2cm 12.8cm 20.1cm 25.5cm,clip=true,width=0.9\textwidth]{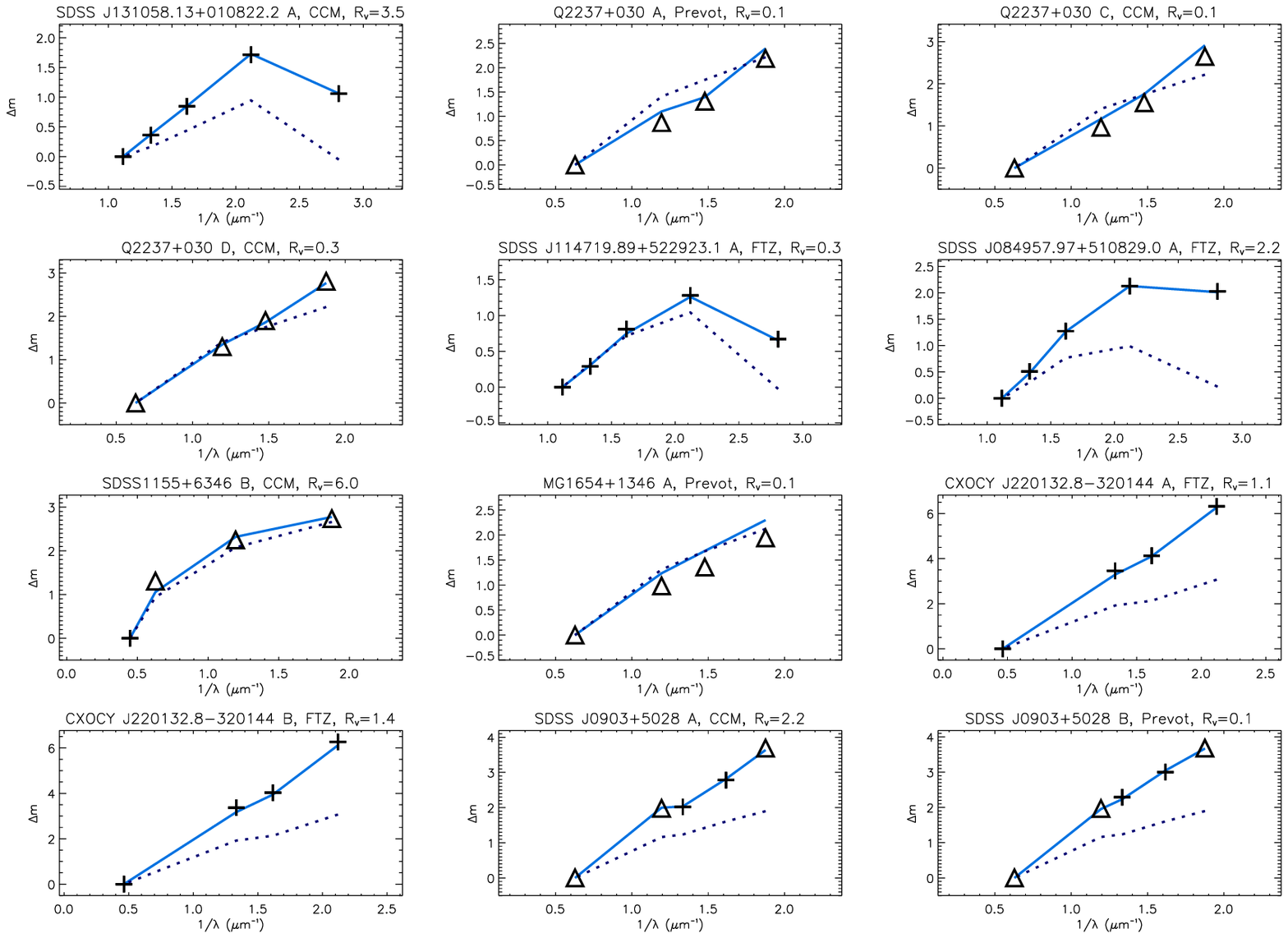}
  \includegraphics[bb=2cm 12.8cm 20.1cm 25.5cm,clip=true,width=0.9\textwidth]{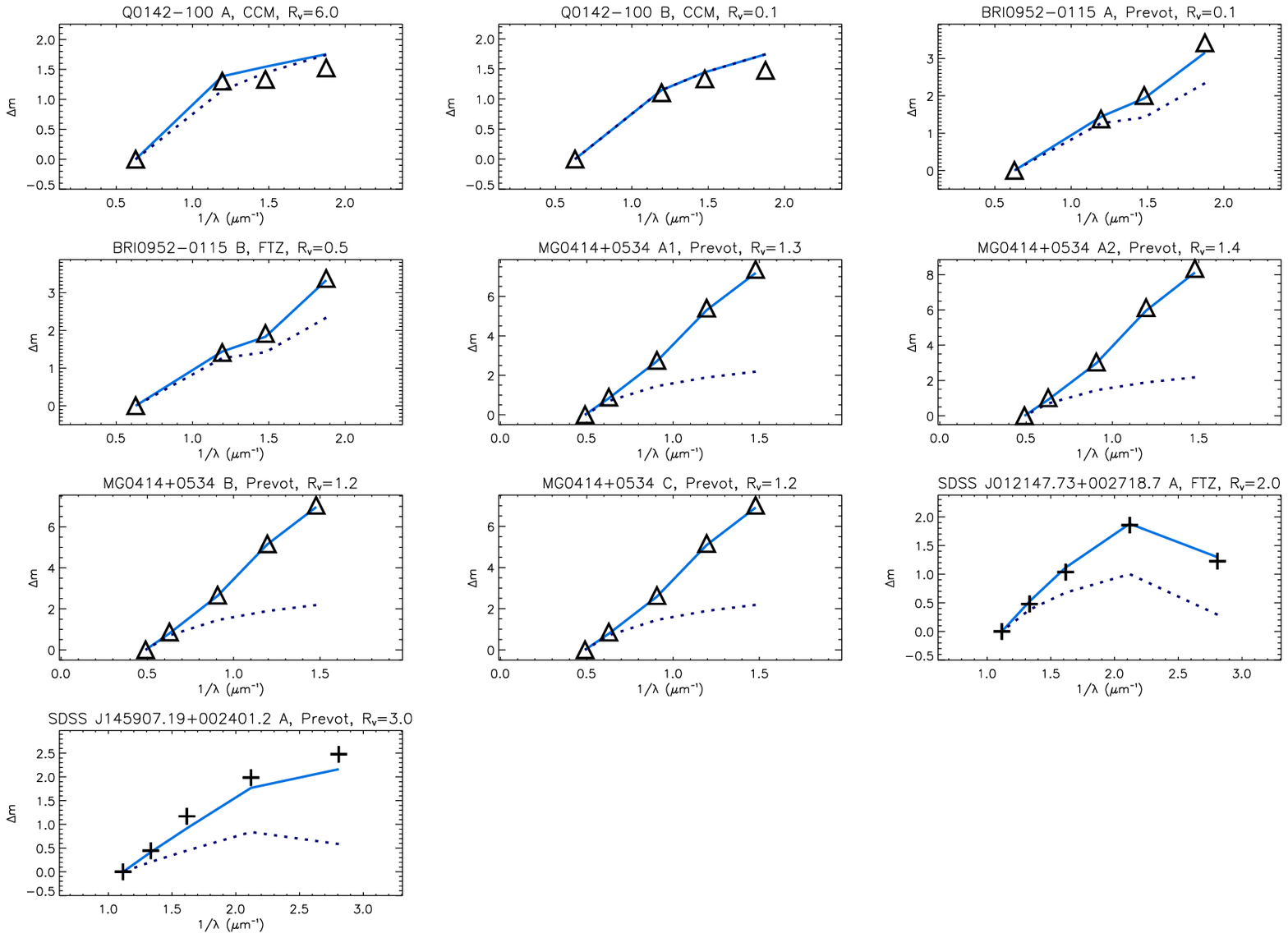}
  \caption{Comparison of the measured magnitudes (black signs) with
  the fitted magnitudes (light blue solid line) for the images that
  could be fitted using the dust laws. The dark blue dotted line shows
  the expected magnitudes without dust. Note that sometimes only one
  line is visible when the best fit is with little or no
  extinction. The ground-based measurements are marked with plus signs
  while the HST observations are marked with triangles. All magnitudes
  are normalised so that the magnitude at the longest wavelength is
  zero.}
  \label{fig:fits}
\end{figure*}

In Figure~\ref{fig:fits}, observed colours are plotted together with
the best synthetic dust colours for each of the quasar images, that
could be fitted by dust (i.e. the images in Table~\ref{tab:fits}). In
the figure, comparison is made with the synthetic magnitude at the
longest wavelength arbitrarily adjusted to match the observations.
However, this was not the case in the actual fitting procedure, which
involves only differences in magnitudes.
In all of these cases, the fitted magnitudes agree very well with
observed values.


\subsection{The badly fitted quasar images}

\begin{figure*}
  \centering
  \includegraphics[bb=2cm 12.8cm 20.1cm 25.5cm,clip=true,width=\textwidth]{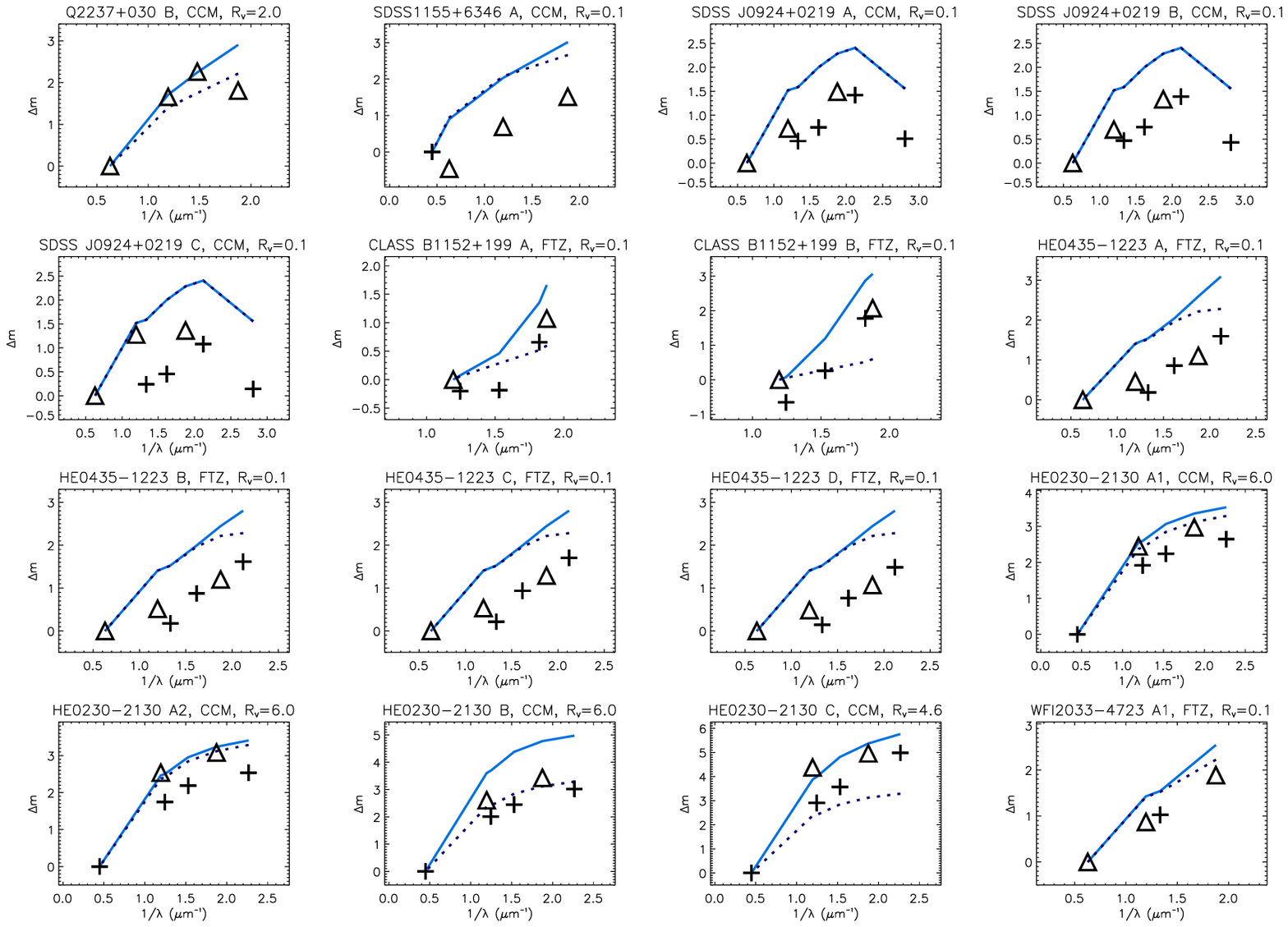}
  \includegraphics[bb=2cm 15.8cm 20.1cm 25.5cm,clip=true,width=\textwidth]{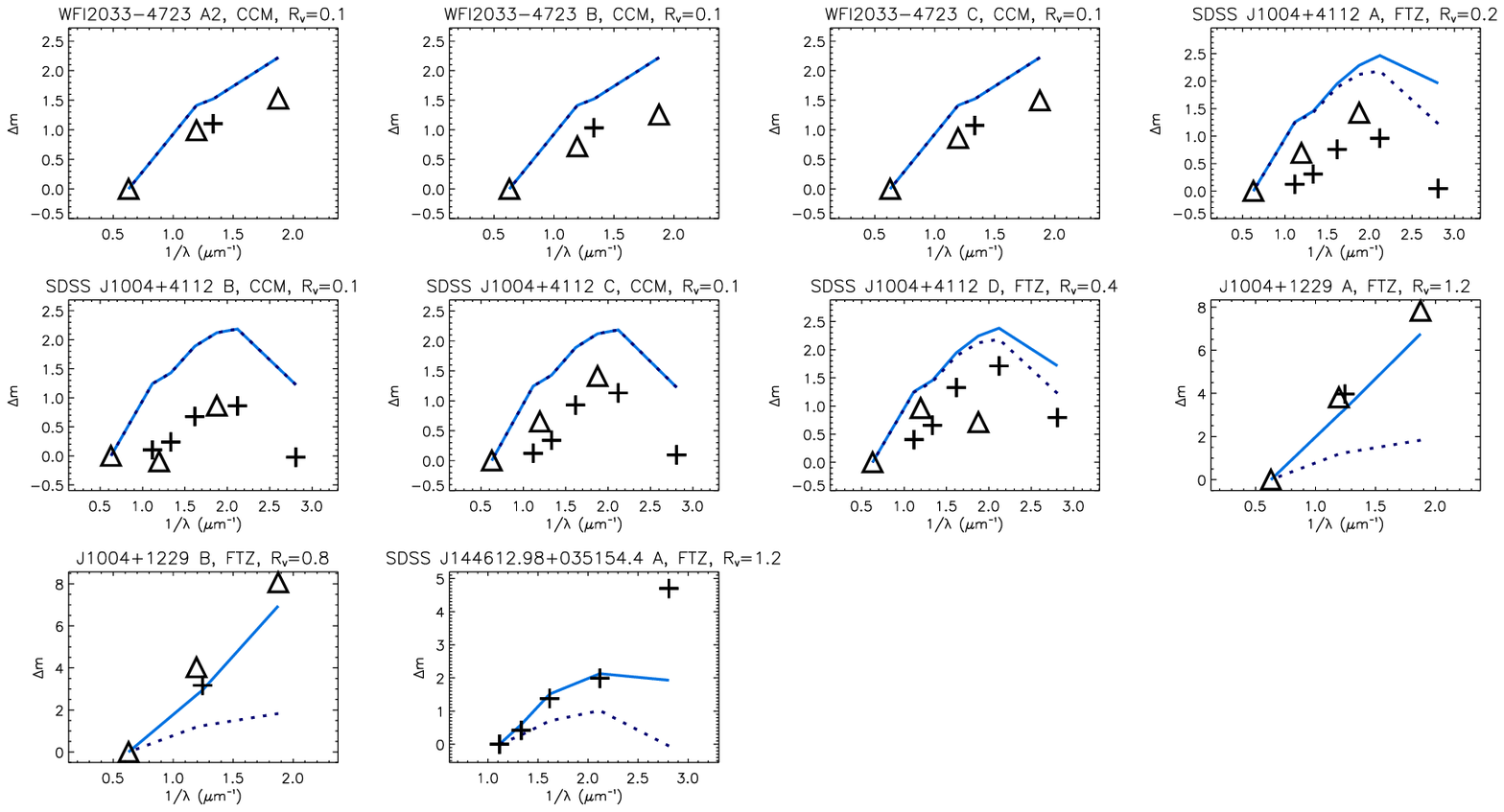}
  \caption{Comparison of the measured magnitudes (black signs) with
  the fitted magnitudes (light blue solid line) for the images that
  were badly fitted using the dust laws. The dark blue dotted line
  shows the expected magnitudes for no extinction. Note that sometimes
  only one line is visible when the best fit is with little or no
  extinction. The ground-based measurements are marked with plus signs
  while the HST observations are marked with triangles. All magnitudes
  are normalised so that the magnitude at the longest wavelength is
  zero.}
  \label{fig:badfits}
\end{figure*}

In Figure~\ref{fig:badfits}, we show how well the best extinction law
agrees with observed magnitudes for each of the quasar images that
could not be fitted with regular dust extinction.

A possible explanation for the difficulty in fitting some of the
quasar images with the reddened template is that in some cases,
observations of the same image in different filters were made at
different times. If the object is variable, and has changed
significantly between the times of observations for the different
filter sets (e.g. HST and ground-based), this could lead to erroneous
observed colours, which cannot be correctly fitted by dust. One
possible example is {\sdsstusenfyra} where the ground-based and HST
observations appear to be shifted with respect to each other (see
Figure~\ref{fig:badfits}). Therefore, in the cases where we have
enough data points, we separate the HST and ground-based observations
to see if this will provide a good fit. There are {\noimground} quasar
images that have observations in a minimum of four different filters,
which are either ground-based or space-based. When applying our
analysis to these data, we find that two of the images of {\hetwo} can
be described by dust extinction. The best-fit values are provided in
Table~\ref{tab:likely}.
\begin{table*}
\caption{Additional best fits of $R_V$ and $E(B-V)$ together with the
1$\sigma$ uncertainties for quasar images which are likely to be dust
extincted. The first three quasar images are badly described when
using both HST and ground-based images, but are possible to fit with
dust when using only the ground-based measurements. The last image is
one where the $u$ band magnitude seems to be deviant from the other
magnitudes and we have made a fit without it.}
\label{tab:likely}
\centering
\begin{tabular}{l c c c c c l c c}
\hline \hline
Quasar & Image & $z_d$ & d (kpc) & $R_V$ & $E(B-V)$ & Law & Prob & Comment\\
\hline
HE0230-2130               &A2     & 0.52  &  6.1  &3.2 (1.5-5.4)&0.2 (0.1-0.2)&CCM              &   13 \%  &  ground \\
HE0230-2130               &A2     & 0.52  &  6.1  &3.4 (1.8-6.0)&0.2 (0.1-0.2)&Fitzpatrick      &   12 \%  &  ground \\
HE0230-2130               &A2     & 0.52  &  6.1  &4.0 (1.6-6.0)&0.2 (0.1-0.3)&SMC              &   15 \%  &  ground \\
HE0230-2130               &A2     & 2.16  &  6.1  &0.1 (0.1-3.0)&0.1 (0.1-0.1)&Host CCM         &    1 \%  &  ground \\
HE0230-2130               &A2     & 2.16  &  6.1  &1.7 (0.4-2.8)&0.1 (0.0-0.1)&Host Fitzpatrick  &    4 \%  &  ground \\
HE0230-2130               &A2     & 2.16  &  6.1  &6.0 (0.1-6.0)&0.1 (0.1-0.1)&Host SMC         &    4 \%  &  ground \\
\hline
HE0230-2130               &B      & 0.52  &  8.0  &2.6 (0.5-4.2)&0.3 (0.2-0.3)&CCM              &    1 \%  &  ground \\
HE0230-2130               &B      & 0.52  &  8.0  &2.9 (0.7-5.7)&0.3 (0.1-0.3)&Fitzpatrick      &    1 \%  &  ground \\
HE0230-2130               &B      & 0.52  &  8.0  &3.7 (1.4-6.0)&0.3 (0.2-0.3)&SMC              &    1 \%  &  ground \\
HE0230-2130               &B      & 2.16  &  8.0  &6.0 (0.1-6.0)&0.1 (0.1-0.1)&Host SMC         &    1 \%  &  ground \\
\hline
SDSS J144612.98+035154.4  &A      & 1.51  &    -  &2.0 (0.1-2.0)&0.2 (0.1-0.3)&CCM              &   26 \%  &    no u \\
SDSS J144612.98+035154.4  &A      & 1.51  &    -  &2.0 (1.4-2.0)&0.2 (0.1-0.2)&Fitzpatrick      &   32 \%  &    no u \\
SDSS J144612.98+035154.4  &A      & 1.95  &    -  &1.1 (0.9-1.5)&0.1 (0.1-0.1)&Host Fitzpatrick  &    9 \%  &    no u \\
\hline
\hline
\end{tabular}
\end{table*}

When calculating the errors of the synthetic colours, we defined
colour outliers as quasars with a colour differing by more than three
standard deviations from the mean. This led to the exclusion of 12\%
of the quasars in our reference set. Thus we would expect about 2.5 of
our 21 quasars to be colour outliers, if our sample is as homogeneous
as the SDSS sample. Some of the badly-fitted quasar data could
correspond to so-called peculiar quasars, differing significantly from
the template in one or several filters.

If a deviation is observed in only one filter, it could be explained
by a strong absorption or emission feature, which is redshifted into
the wavelength region of the filter of observation. Some badly-fitted
images, such as SDSS J144612.98+035154.4, have colours that agree well
with synthetic dust colours, apart from one band (see
Figure~\ref{fig:badfits}). For these, we make the assumption that the
diverging magnitude is erroneous and redo the fit without. Because
observations in at least four filters are required, there is only one
case of interest, {\sdssettfyrafyra}. The best-fit values for this
system, when removing the $u$ band magnitude are provided in
Table~\ref{tab:likely}.


\subsection{Global results}

\begin{figure}
  \centering
  \includegraphics[width=0.4\textwidth]{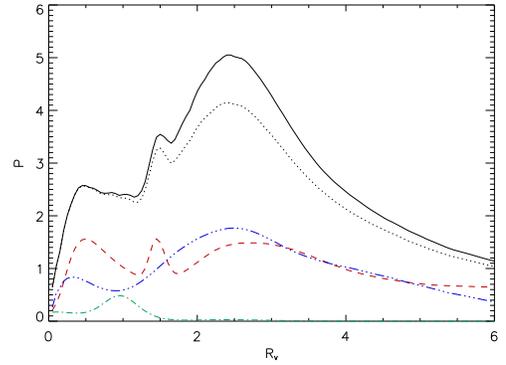}
  \caption{The co-added probability distributions for the fits with
  intervening dust following the {\fitzpatrick} extinction law,
  weighted by the probability of the fit. The dotted line shows the
  added probability when using the well fitted quasar images in
  Table~\ref{tab:fits} and the solid line when also the objects in
  Table~\ref{tab:likely} are included. The added probabilities for the
  subset of early-type galaxies (dashed line), late-type galaxies
  (dash-dot) and starburst/star-forming galaxies (dash-dot-dot-dot) are
  also included in the figure.}
 \label{fig:ideo}
\end{figure}

\begin{figure*}
  \centering
  \includegraphics[width=0.4\textwidth]{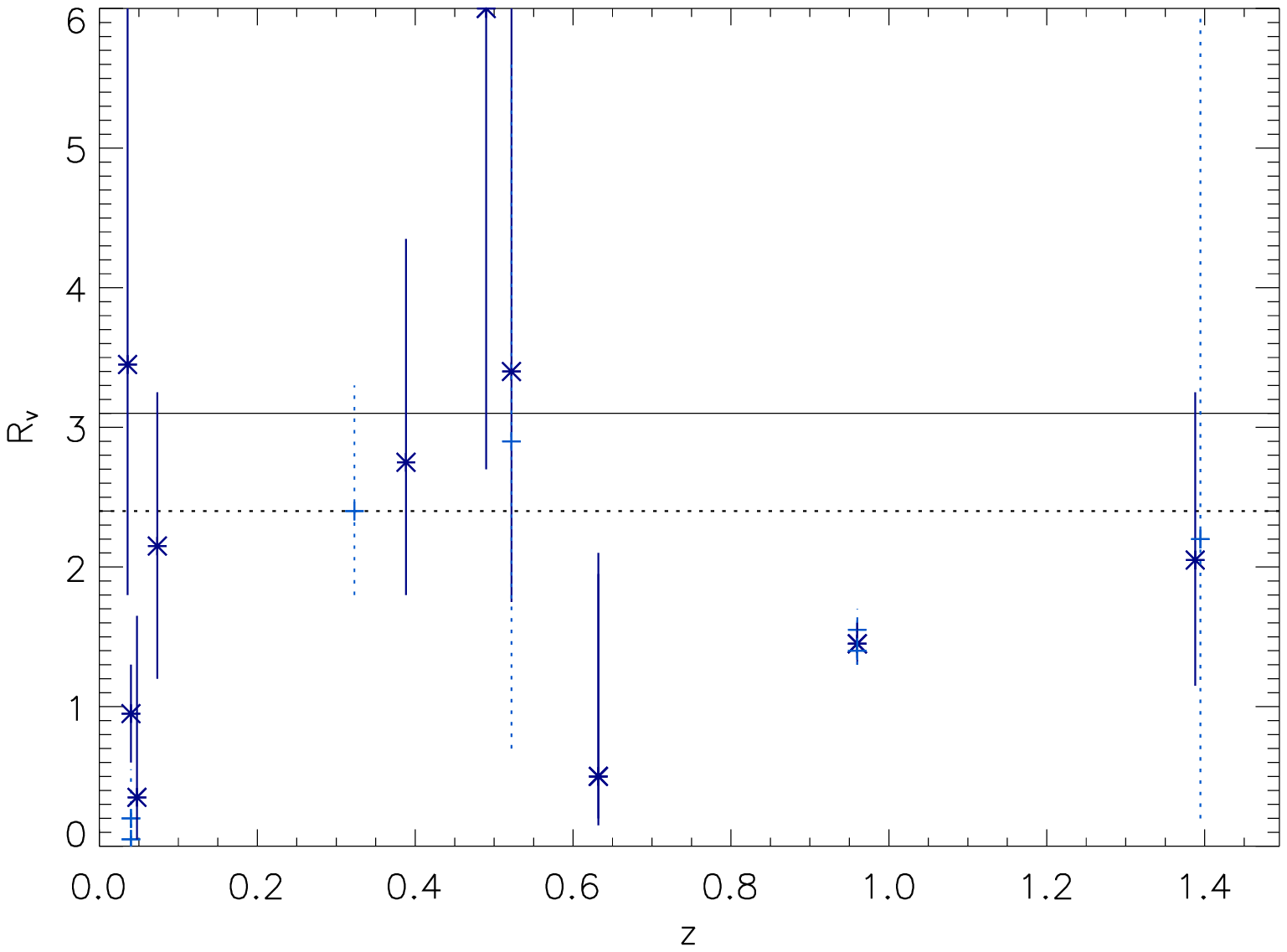}
  \includegraphics[width=0.4\textwidth]{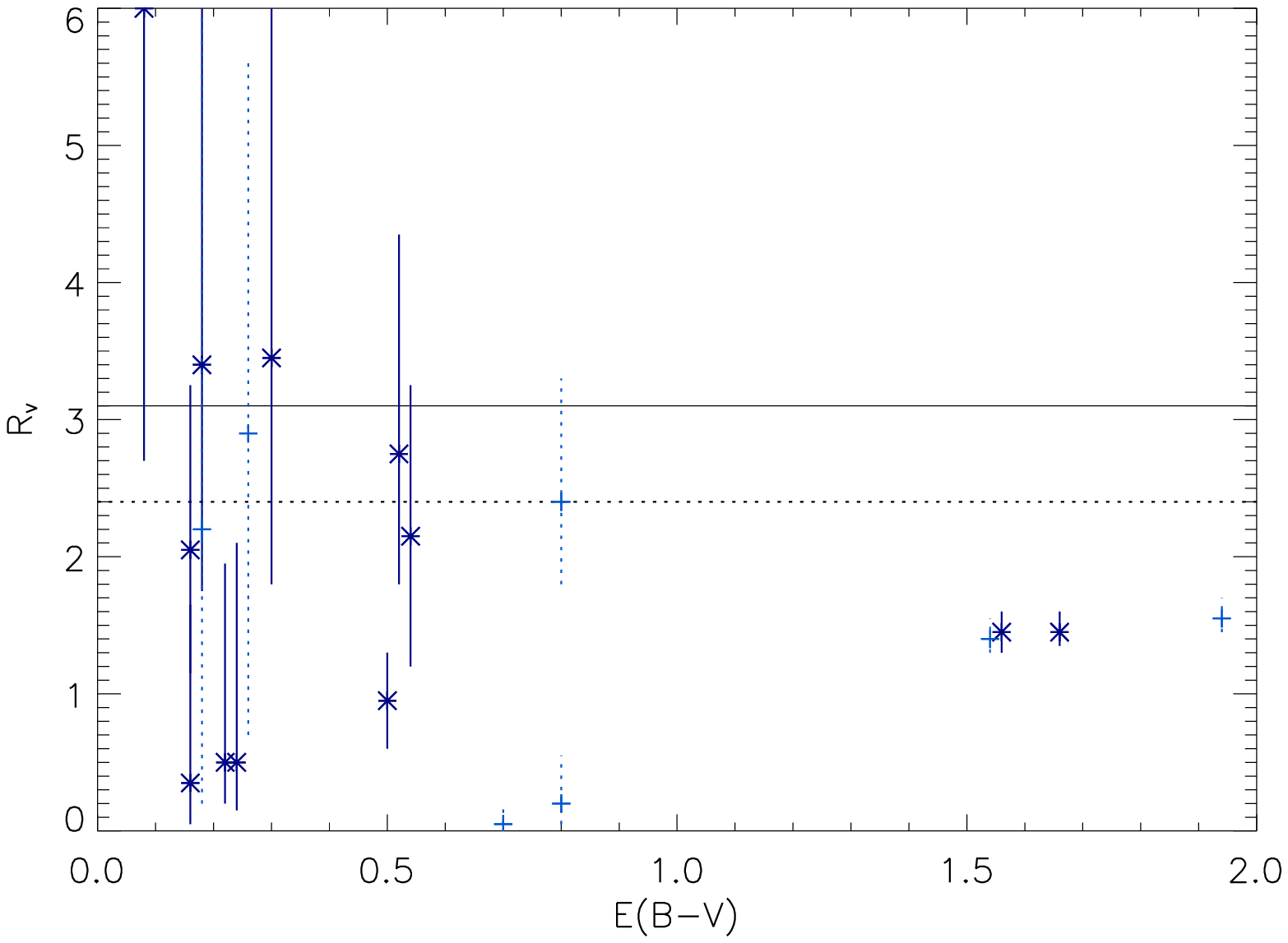}
  \caption{The fitted values of $R_V$ with the 1$\sigma$ errorbars for
  all images that could be fitted using the {\fitzpatrick}
  parameterisation of dust in an intervening galaxy. The first panel
  shows $R_V$ as a function of dust redshift and the second as a
  function of colour excess. In the first panel, the results from the
  different images of the same quasar will be plotted on top of each
  other since they all have the same redshift. In the second panel,
  the results of different images will be plotted in different places
  along the x axis since $E(B-V)$ differs at least slightly between
  the images. The fitted values which have a larger probability than
  10\% (see Table~\ref{tab:fits}) are plotted with stars and dark blue
  solid errorbars while the less reliable ones, with a smaller
  probability than 10\%, are plotted with plus signs and dotted light
  blue errorbars. The most probable value of $R_V$ found when
  co-adding the probability distributions for the fits (see
  Figure~\ref{fig:ideo}) is indicated with a dotted line at $R_V = $
  {\ideorv}.}
  \label{fig:rv}
\end{figure*}

To understand how the fitted values of $R_V$ are distributed, we
assume that all quasars are affected by dust extinction in the
intervening galaxy, following the {\fitzpatrick} law. For each image,
we calculate a probability distribution for $R_V$, by locating the
maximum $\chi^2$ probability for each $R_V$, when $E(B-V)$ is set
free. The maximum of the distribution will have the same value, as the
probability for the {\fitzpatrick} law, in Table~\ref{tab:fits}. The
probability distributions for all images that could be fitted with
dust, are then added together. In this way, we take into account the
possibility of non-Gaussian distributions, and the result is weighted
by the probability of the fit. We exclude images for which our fits
are consistent with almost no extinction because the $R_V$ dependence
is weak then.
The result is presented in Figure~\ref{fig:ideo}, where it is evident
that the inclusion of the objects in Table~\ref{tab:likely}, does not
affect the shape of the curve significantly.
Using the larger set, we found a most probable value of $R_V$ of
{\ideorv}, with a FWHM of {\ideorvfwhm}. This value is slightly lower
than the Galactic value of 3.1, but well within the range due to the
large spread of fitted values. In Figure~\ref{fig:rv}, the best-fit
values of $R_V$ are plotted as a function of redshift and $E(B-V)$.
There are no clear trends in the dependence with these parameters on
$R_V$.

To test the possibility of a dependence of galaxy type for $R_V$, we
divide Figure~\ref{fig:ideo} into three plots, one for early-type
galaxies, one for late-type galaxies, and one for
starburst/star-forming galaxies. These are shown in
Figure~\ref{fig:ideo}. The plot for the late-type galaxies have less
features than the one for all galaxies, but the most probable $R_V$
remains similar. The plots for the early-type galaxies and the
starburst galaxies are more different. However, these are constructed
using only four and three images, respectively. It would be
interesting to repeat the same study for a larger number of galaxies
in the future.

We note that since a large part of the galaxies studied are lensing
galaxies, the sample will not necessarily be representative of the
universal galaxy population, because strong-lensing surveys are more
efficient in finding massive galaxies.


\subsection{Specific systems}

Several of the quasar-galaxy systems that we study, have multiple
images. Because the multiple images have the same intrinsic colours,
the difference in colour between the images should be explained by
dust extinction, or possibly by microlensing or time-variability
effects. In this section, we compare fits for different images of the
same quasar.
We also compare with other dust property estimates of the same object
from the literature. In \citet{ostman06}, some of the objects here
were investigated with a similar method, that has now been
improved. The most important improvement is that, while in
\citet{ostman06}, we compared all magnitudes with the value in the $u$
band in the expression for the $\chi^2$, we here optimise the choice
of colours to maximise the probability of detecting a potential dust
extinction (see Section~\ref{sec:method}). Another improvement is that
we include the observational errors, in addition to the template errors,
in the covariance matrix.
Some of the systems that we have studied have also been investigated
using the differential method by other authors
\citep{falco99,ardis,toft00}. When we state their preferred values of
$R_V$ in the text below followed by an interval within parenthesis,
this is the error interval of $R_V$ provided in that paper.



\paragraph{SDSS J131058.13+010822.2}

This quasar-galaxy system was found in \citet{ostman06}, by matching
the coordinates of quasars with those of galaxies. The colours of the
quasar indicate that it is most likely affected by dust extinction. In
that paper, it was found to have $R_V=3.4 (1.7-5.7)$ using the
{\cardelli} parameterisation, and $R_V=3.4 (2.4-4.7)$ using the
{\fitzpatrick} parameterisation. This agrees well with what was found
in this analysis, where the preferred value of $R_V$ was 3.5.

According to the New York University Value-Added Galaxy Catalog
\citep{nyuvagc}, the foreground galaxy is a star-forming galaxy at
$z=0.04$. Fitting the quasar colours with starburst-like extinction
according to the {\calzetti} law, we find a best fit $E(B-V)$ of 0.3
(0.2-0.4), where the $\chi^2$ provides a probability of 99\%.


\paragraph{Q2237+030} 

There are four images of this quasar, three of them can be fitted with
dust extinction in the intervening galaxy, all with unrealistically
low values of $R_V$ and an $E(B-V)$ between 0.5 and 1.2, depending
upon which image is studied and the parameterisation that is used. The
extremely low preferred values of $R_V$ for image A and C, are
difficult to justify physically due to the strange appearance of the
extinction laws for low $R_V$ (e.g. see the {\cardelli}
parameterisation with $R_V=0.5$ in Figure~\ref{fig:extlaws}). Image D
has a larger $R_V$ value, with a higher chi-square probability. Image
B, which has one deviating band, could not be fitted by dust
extinction.

The extinction properties of this system have been measured by
\citet{falco99}, using the differential method. They obtained a high
value of $R_V$ of $5.29 (4.47-6.11)$, which is inconsistent with our
result. \citet{ardis} used the same technique and derived values of
$R_V$ between 2.9 and 3.1, with errors of the order of 1.5. In our
analysis, values below 1.3, were obtained.

We note that the colours of the object have been shown to vary with
time \citep{moreau05} and therefore there must be an effect, in
addition to or instead of dust extinction, affecting the colours, for
example microlensing.
This would explain why we derive best-fit values for $R_V$ that are
unrealistically low, and why completely different values have been
obtained using other data sets.


\paragraph{SDSS J114719.89+522923.1}  

This case of a quasar shining through a galaxy, was discovered by
\citet{ostman06}. However, no $R_V$ could be determined.
\citet{konig06} have since detected absorption features from the
intervening galaxy in the quasar spectrum strengthening the assumption
that the colours of this quasar are affected by dust absorption in the
intervening galaxy. In this paper, we found that the colours are
best-fitted by dust in the intervening galaxy, with a low $R_V$.

According to the New York University Value-Added Galaxy Catalog
\citep{nyuvagc}, the foreground galaxy is a starburst galaxy at
$z=0.05$. We find that Galactic-like dust describes the observed
colours better than typical dust in starburst galaxies. Using the
{\calzetti} law for starburst extinction, we find a best-fit value of
$E(B-V)$ of 0.2, with a $\chi^2$ probability of 12\%.


\paragraph{SDSS J084957.97+510829.0}  

This quasar with a foreground galaxy was discovered by
\citet{ostman06}. It was found to have a measured value of $R_V$ of
$1.7 (0.7-2.9)$ using the {\cardelli} parameterisation, and of $2.2
(1.5-2.9)$ using the {\fitzpatrick} parameterisation. This agrees well
with our preferred values of $R_V$ for this system, of approximately
2.2-2.3.

According to the New York University Value-Added Galaxy Catalog
\citep{nyuvagc}, the foreground galaxy is a starburst galaxy at
$z=0.07$. We found that the observed colours of the quasar were
reproduced more effectively using the {\fitzpatrick} extinction curve,
than using an extinction curve that should be valid for starburst
galaxies. However, the fit with starburst-like extinction according to
the {\calzetti} law was also good. The best fit with $E(B-V)=0.5$
provided a $\chi^2$ probability of 44\%.


\paragraph{SDSS1155+6346}

This quasar has two images. For one image, we were unable to explain
the measured colours by dust extinction, and for the other image,
there is only a small probability that the best-fit dust scenario is
the true explanation of the colours. The magnitudes of image B could
be fairly well-described by dust, if the observational errors were
larger. Image A, on the other hand, has a deviating K band
magnitude. This is one case where one could suspect problems with the
ground-based observations. It would be interesting to observe this
object further to investigate if the deviation is a result of the
observations, or if it is an interesting feature in the spectrum.


\paragraph{MG1654+1346}  

The colours of MG1654+1346 could be explained by SMC-like dust in the
intervening galaxy, with a low $R_V$. However, the $\chi^2$ of the fit
is high.


\paragraph{CXOCY J220132.8-320144}

There are two images of CXOCY J220132.8-320144, both of which could be
explained by dust, in particular at the host galaxy with low values of
$R_V$.


\paragraph{SDSS J0903+5028}

This object was studied by \citet{ostman06}, using a different data
set. The {\cardelli} parameterisation gave $R_V=0 (0-0.7)$, and the
{\fitzpatrick} parameterisation $R_V=0.9 (0.7-1.2)$, for the image we
will refer to as image A. However, we did not manage to fit this image
with intervening extinction here, and we have therefore no $R_V$ with
which to compare. Instead we found that it could be described by
extinction in the host galaxy with an $R_V$ of approximately
2.3-2.8. Image B, however, can be reproduced by assuming that there is
dust either in the intervening galaxy or in the host galaxy.


\paragraph{SDSS J0924+0219}

There are three images of SDSS J0924+0219. The colours within these
images cannot be fitted by the extinction laws used in our analysis.


\paragraph{CLASS B1152+199}

Using the differential method, both \citet{toft00} and \cite{ardis}
have measured $R_V$. They obtained $R_V=1.3-2.1$ and $R_V=2.1
(2.0-2.2)$, respectively. In our analysis, we have not managed to find
a good fit to the observed magnitudes for any of the two images.


\paragraph{HE0435-1223} 

There are four images of this quasar. None of these images have
measured colours that could be fitted using our extinction laws.


\paragraph{Q0142-100}

Using the differential method, \citet{falco99} and \citet{ardis} have
measured $R_V$ for this system. They obtained $3.11 (2.11-4.11)$ and
$R_V = 4.7 (4.0-5.4)$, respectively. We find that there is little
extinction in this system, causing unreliable $R_V$ determinations.


\paragraph{HE0230-2130}

None of the four images of HE0230-2130 could be reproduced by taking
into account the effects of dust, using both the HST, and the
ground-based observations. However, if we use only the ground-based
observations, some of the images can be explained by dust, in
particular image A2. This is one of few images for which we measure a
higher $R_V$, than for the Milky Way, 3.2 for the {\cardelli}
parameterisation and 3.4 for the {\fitzpatrick} parameterisation.


\paragraph{BRI0952-0115}

\citet{falco99} fitted a value of $3.10 (2.10-4.10)$ for $R_V$, using
the differential method. With intervening extinction, we found lower
values of $R_V$.


\paragraph{WFI2033-4723}

There are four images of this quasar. None of these images have
colours that could be reproduced well by our extinction laws.


\paragraph{SDSS J1004+4112}  

There are four images of this quasar. The observations could not be
fit by assuming dust extinction for any of the images. The
ground-based observations appear to follow a smooth curve, but the HST
observations appear to be more scattered. However, using only the
ground-based observations, we were unable to obtain a good fit, using
the extinction laws used in this paper.


\paragraph{J1004+1229}

There are two images of this quasar. We were unable to reproduce the
colours of either image, using our dust extinction laws.


\paragraph{MG0414+0534}

This is an interesting case of a quasar that is heavily
reddened. There have been claims both that the reddening is caused by
dust in the lens galaxy, and in the host galaxy. \citet{lawrence}
reason that the most likely cause is dust extinction in the lens
galaxy. They argue that (1) the observed spectrum can be
well-explained by dust at intervening redshifts\footnote{They used the
wrong lensing redshift, but according to our analysis with colour
comparisons the claim is still valid for the correct redshift.}, while
extinction in the host galaxy is unable to reproduce the spectrum for
Galactic extinction and barely for SMC-like extinction; (2) the
separation between the light paths for the different images at the
host is too small to explain the difference in colours between the
images; in contrast, a separation at the lensing galaxy would be able
to reproduce the colour difference better; and (3) two of the reddest
quasars that are known, MG 0414+0534 and MG 1131+0456, both have an
intervening galaxy, and it is unlikely that both have a foreground
galaxy, and that the foreground galaxy is not responsible for the
extremely red colours. However, for the case of MG 1131+0456,
\citet{kochanek00} concluded that dust is not responsible for the red
colour, at either redshift. They proposed instead that stellar
emission from the host galaxy, could explain the red
colours. \citet{tonry} suggest that the red colours of MG0414+0534 are
caused by dust extinction in the host.  Their arguments are that (1)
the visible arc of the host galaxy is bluer than the quasar images and
that the extinction must be uniform over a large region, because there
is little difference in extinction between quasar images, and no
notable difference in colour along the arc; and (2) that the colour
and surface brightness of the lens galaxy agrees well with that of a
passively-evolving early-type galaxy on the Fundamental Plane
\citep{keeton}.

\begin{figure*}
  \centering
  \includegraphics[width=0.4\textwidth]{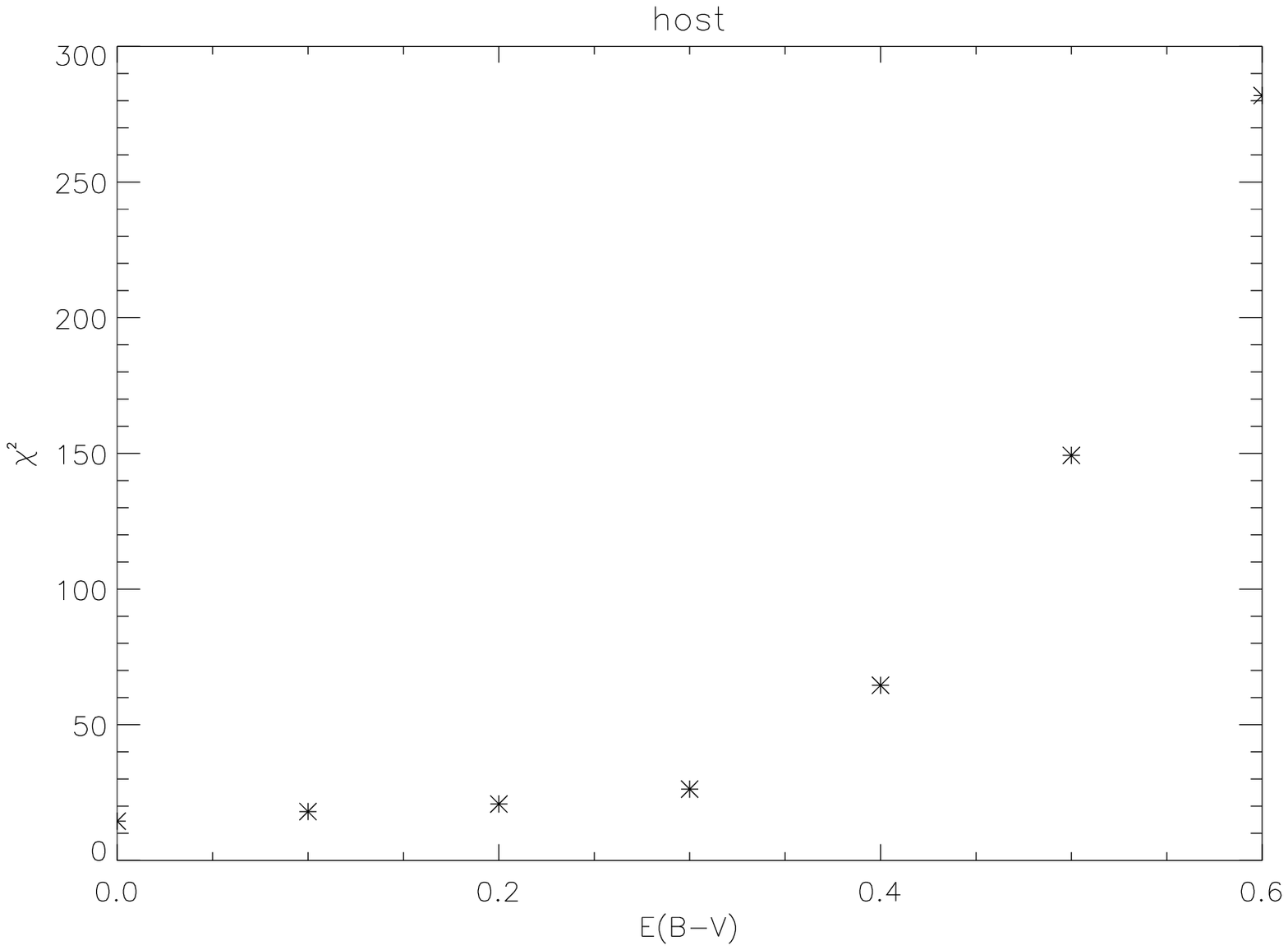}
  \includegraphics[width=0.4\textwidth]{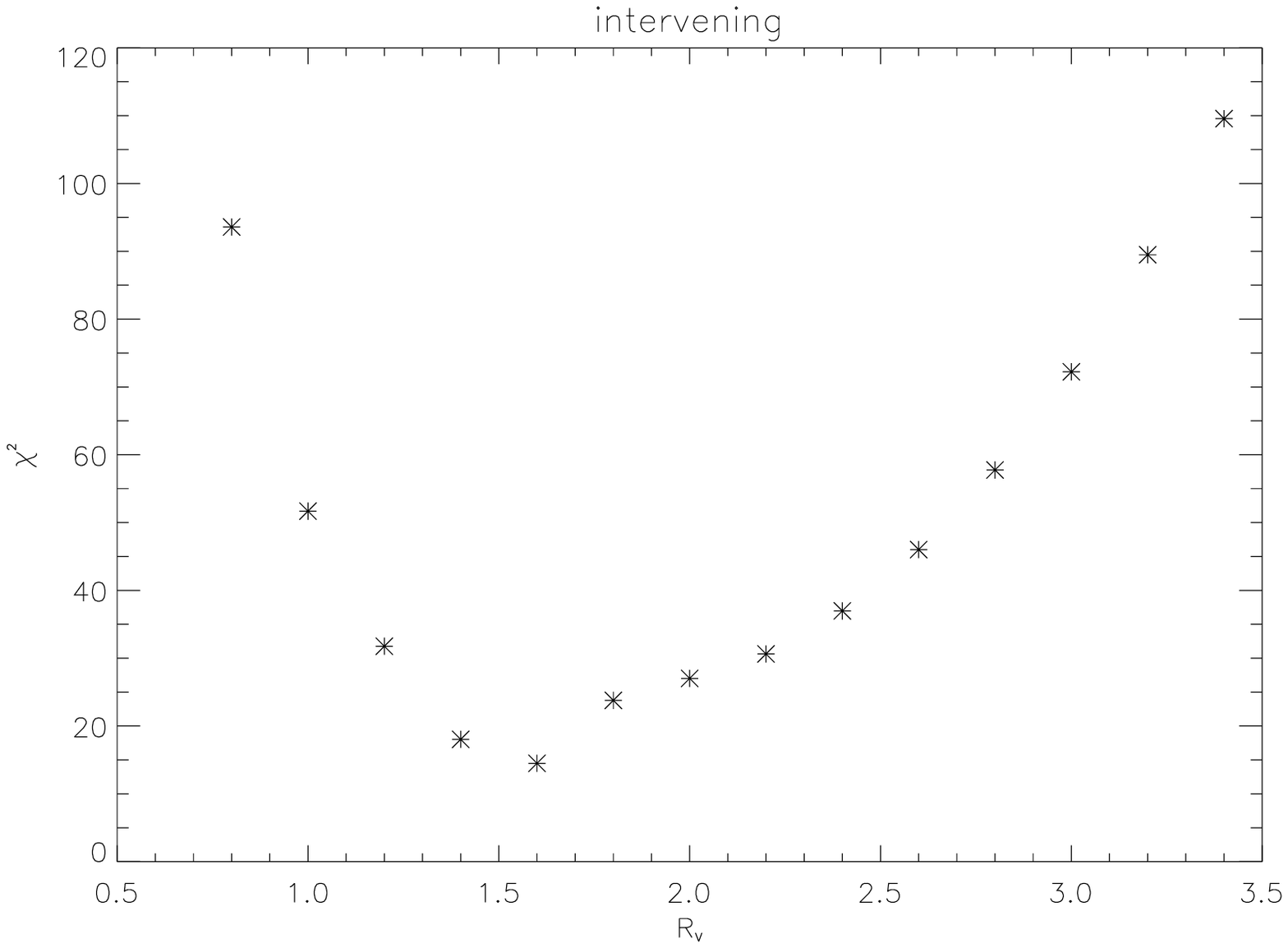}
  \caption{The $\chi^2$ values for different $E(B-V)$ in the host
  galaxy (first panel) and $R_V$ in the intervening galaxy (second
  panel) for {\mgfour}.}
  \label{fig:multi}
\end{figure*}

When fitting for dust extinction, we found that the colours of the
quasar could be explained by presence of dust in the lens galaxy, but
not by dust in the host galaxy alone. All images gave reasonably
similar values of $R_V$, indicating that the foreground galaxy has a
homogeneous dust content. Since there are strong claims of dust
extinction in the host galaxy, we decided to measure the likelihood of
dust extinction in both galaxies. We fitted for dust extinction at
both redshifts with one $R_V$ for each redshift, a value of $E(B-V)$
for the host extinction, and a $E(B-V)$ value for each line of sight
through the foreground galaxy. We found that the best fit was given by
no extinction in the host galaxy [$E(B-V)=0$], and severe extinction
in the foreground galaxy [$R_V=1.5$, and $E(B-V)=1.6, 1.9, 1.5, 1.5$
for the different sight lines through the galaxy]. 
Marginalising over $E(B-V)$ for the host galaxy, we find that the
$\chi^2$ is almost as good for small extinctions $E(B-V) < 0.3$ as for
the best fit which had $E(B-V)=0$. For higher values of $E(B-V)$,
however, the $\chi^2$ is much worse (see the first panel of
Figure~\ref{fig:multi}).
From our analysis, it appears that the bulk of the extinction occurs
in the intervening galaxy, but we cannot exclude an additional small
extinction in the host.

The best fit for the intervening galaxy is $R_V \sim 1.5$, regardless
of whether we include extinction in the host or not. In the second
panel of Figure~\ref{fig:multi}, there is a marginalisation over $R_V$
in the intervening galaxy, showing how the $\chi^2$ changes for
different $R_V$. \citet{falco99} found a similar value of $R_V$, $1.47
(1.32-1.62)$. \citet{ardis} found values ranging between 0.4 and 3.8,
depending on the images that they compared.


\paragraph{SDSS J012147.73+002718.7}

The best-fit values for $R_V$ from \citet{ostman06} for this quasar
system (3.3 for the {\cardelli} parameterisation and 1.8 for the
{\fitzpatrick} parameterisation), are within the 1$\sigma$ uncertainty
of our fit.


\paragraph{SDSS J145907.19+002401.2}

This object was also studied by \citet{ostman06}. In that paper, the
{\cardelli} parameterisation provided an $R_V$ value of $0 (0-0.7)$,
and the {\fitzpatrick} parameterisation, an $R_V$ value of $1.8
(0.4-3.5)$. These results are consistent with the values derived in
this paper.


\paragraph{SDSS J144612.98+035154.4}

In \citet{ostman06}, the {\cardelli} parameterisation gave $R_V = 0
(0-0.7)$ and the {\fitzpatrick} parameterisation $R_V=1.3
(1.0-1.6)$. In this paper, we found no dust model that could explain
the colours in a satisfactory way. However, if we ignore the $u$ band
magnitude, the colours could be fitted by intervening dust with an
$R_V$ of 2.0.


\section{Summary and conclusions}
\label{sec:summary}

The reddening of quasars by foreground galaxies has been
investigated. At our disposal, we have had {\nosys} quasar-galaxy
systems with a total of {\noim} quasar images. By looking at the
rest-frame $B-V$, we conclude that the foreground galaxy, close to the
line of sight, must affect the quasar colours. Using redshifted
reddened quasar templates, we fit for $R_V$ and $E(B-V)$ using the
measured colours.

Of the data that we consider, {\noimwell} images could be fitted by
dust extinction. Assuming dust extinction in the intervening galaxy
following the {\fitzpatrick} law, we find a most probable $R_V$ value
of {\ideorv}. This value is lower than the Galactic mean value of
3.1. However, the FWHM of our distribution of $R_V$ was determined to
be {\ideorvfwhm} and thus the Galactic mean value is within the range
of the spread. We note that our sample will not be representative of
the universal galaxy population because the majority of our quasars
were observed by strong lensing surveys, which are particularly
efficient in finding massive galaxies. The large spread in values
indicates that the dust content can vary significantly between
different galaxies. However, we found no strong correlation between
the values of $R_V$ and redshift, colour excess or galaxy type.
In the future, when a larger dataset with superior quality is
available, a similar analysis could be of considerable benefit to, for
example, the use of extinction-corrected supernovae as distance
indicators.
It would not be surprising if a correlation was found, because
metallicity and elemental abundance ratios are expected to affect dust
properties, and both quantities evolve with redshift and galaxy
type. Furthermore, there are at least two major mechanisms that create
dust grains, the first is in AGB stars \citep[e.g.][]{mathis90} and
the second in core-collapse supernovae \citep[e.g.][]{bank03}, which
could give rise to different types of dust.

The uncertainties in the determination of the amount of dust are
large, when taking into account the different possible dust
models. This fact may be of importance for flux anomalies, used as a
probe of cold dark matter substructure \citep{kochanek04}.

Another interesting observation is that when several images of the
same quasar could be fitted by dust, the preferred values of $R_V$
were similar. This could indicate that $R_V$ has little spread within
individual galaxies, but that it varies between galaxies.
A caveat, however, is that for many of our quasars one or several of
the images could not be fitted by the standard reddening laws we
tried, potentially challenging the conclusion of homogeneous dust
properties within individual lensing galaxies, unless the colour
outliers can be attributed to something different from dust.
In total, there were seven quasars in our sample that had several
images, one or several of which could be fitted with dust. Three of
those quasars had similar values of $R_V$, for all images, regardless
of extinction law, two quasars had at least one image that could not
be reproduced by assuming the presence of dust, one quasar had images
where the $R_V$ values did not agree with each other, and one quasar
was consistent with experiencing no dust extinction.

Several systems were found with low values of $R_V$, comparable to
what has been measured from global fits, using Type Ia supernova
data. However, the distribution of $R_V$ for our fits suggests that a
large fraction of the galaxies considered in this study are compatible
with Milky Way dust. Due to the limited number of objects in our
study, as well as the possibility of selection effects and analysis
bias, no strong conclusion can be drawn as to whether the extinction
properties in foreground galaxies, along the line of sight to quasars,
differ or not from extinction in host galaxies of Type Ia supernovae.


\begin{acknowledgements}

The authors would like to thank Vallery Stanishev for useful
discussions and the G\"{o}ran Gustafsson Foundation and the Swedish
Research Council for financial support. EM also acknowledges support
from the Anna-Greta and Holger Crafoord fund.

Funding for the SDSS and SDSS-II has been provided by the Alfred
P. Sloan Foundation, the Participating Institutions, the National
Science Foundation, the U.S. Department of Energy, the National
Aeronautics and Space Administration, the Japanese Monbukagakusho, the
Max Planck Society, and the Higher Education Funding Council for
England. The SDSS is managed by the Astrophysical Research Consortium
for the Participating Institutions. The Participating Institutions are
the American Museum of Natural History, Astrophysical Institute
Potsdam, University of Basel, Cambridge University, Case Western
Reserve University, University of Chicago, Drexel University,
Fermilab, the Institute for Advanced Study, the Japan Participation
Group, Johns Hopkins University, the Joint Institute for Nuclear
Astrophysics, the Kavli Institute for Particle Astrophysics and
Cosmology, the Korean Scientist Group, the Chinese Academy of Sciences
(LAMOST), Los Alamos National Laboratory, the Max-Planck-Institute for
Astronomy (MPIA), the Max-Planck-Institute for Astrophysics (MPA), New
Mexico State University, Ohio State University, University of
Pittsburgh, University of Portsmouth, Princeton University, the United
States Naval Observatory, and the University of Washington.

\end{acknowledgements}


\bibliographystyle{aa}
\bibliography{ms9187}

\end{document}